\documentclass[prx,10pt,onecolumn,groupedaddress,floatfix, showpacs,nofootinbib]{revtex4-2}

\usepackage[sc,osf]{mathpazo}
\usepackage[scaled=0.86]{berasans}  
\usepackage[colorlinks=true, allcolors=blue, urlcolor=blue]{hyperref}  
\usepackage{graphicx} 
\usepackage{amsmath,mathtools,amssymb,amsthm,bm,amsfonts,mathrsfs,bbm} 

\usepackage{xspace}  
\usepackage{pgfplots}
\usepackage{xcolor,colortbl}
\usepackage{array}
\usepackage{bigstrut}
\usepackage{mathrsfs}
\usepackage{dsfont}
\usepackage{quantikz}
\usepackage{multirow}
\usepackage{makecell}

\usepackage{caption}
\usepackage{subcaption}
\usepackage{ragged2e}
\DeclareCaptionJustification{justified}{\justifying}
\usepackage{tabularx}
\newcolumntype{C}{>{\centering\arraybackslash}X}

\usepackage{lipsum}

\usepackage{verbatim}

\usepackage{comment}

\newcommand{\tr}{\text{tr}}

\newcommand{\bracket}[3]{\langle #1|#2|#3 \rangle}

\newcommand{\be}{\begin{equation}}
\newcommand{\ee}{\end{equation}}
\newcommand{\bea}{\begin{eqnarray}}
\newcommand{\eea}{\end{eqnarray}}
\newcommand{\bes}{\begin{equation*}}
\newcommand{\ees}{\end{equation*}}
\newcommand{\beas}{\begin{eqnarray*}}
\newcommand{\eeas}{\end{eqnarray*}}

\renewcommand{\ket}[1]{|#1\rangle}
\newcommand{\ketbra}[1]{\ket{#1}\!\bra{#1}}
\renewcommand{\bra}[1]{\langle#1|}

\newcommand{\I}{\mathds{1}}

\renewcommand{\H}{\mathcal{H}}

\def\tr{\mathrm{tr}}

\begin{document}

  



\title{ Purification of Noisy Measurements and Faithful Distillation of  Entanglement  }


\author{Jaemin Kim}
\email{woals6584@kaist.ac.kr} 

\author{Jiyoung Yun}
\email{jiyoungyun@kaist.ac.kr}

\author{Joonwoo Bae}
\email{joonwoo.bae@kaist.ac.kr}

\affiliation{School of Electrical Engineering, Korea Advanced Institute of Science and Technology (KAIST), $291$ Daehak-ro, Yuseong-gu, Daejeon $34141$, Republic of Korea }


\begin{abstract}
We consider entanglement distillation with noisy operations in which quantum measurements that constitute a general quantum operation are particularly noisy. We present a protocol for purifying noisy measurements and show that imperfect local operations can distill entanglement. The protocol works for arbitrary noisy measurements in general and is cost-effective and resource-efficient with single additional qubit per party to resolve the distillation of entanglement. The purification protocol is feasible with currently available quantum technologies and readily applied to entanglement applications.
 \end{abstract}

\maketitle

\section{Introduction}

Entanglement is a resource that enables efficient information processing \cite{Horodecki2009, GUHNE20091} such as quantum computation \cite{PhysRevLett.86.5188, PhysRevA.68.022312, Briegel:2009aa} and secure communication \cite{PhysRevLett.92.217903, PhysRevLett.94.020501, doi:10.1142/S0219749908003256, PhysRevLett.98.230501}. Manipulating shared entangled states by local operations and classical communication (LOCC) is a key to entanglement applications. In particular, an entangled bit (ebit),  
\bea
|\phi^{+}\rangle^{(AB)} = \frac{1}{\sqrt{2}} (|00\rangle^{(AB)} +|11\rangle^{(AB)})  \label{eq:ebit}
\eea 
is the unit of entanglement \cite{PhysRevA.53.2046, PhysRevLett.76.722, Bennett_1996, PhysRevLett.77.2818, PhysRevA.65.032321}.


Distilling ebits from noisy entanglement by LOCC, firstly proposed in Refs. \cite{PhysRevLett.76.722, PhysRevLett.77.2818}, makes it possible for two parties far apart to share entangled states. Once ebits are distilled, parties can extend a distance through quantum repeaters \cite{PhysRevLett.81.5932}. Entanglement distillation  \cite{PhysRevLett.76.722, PhysRevLett.77.2818} has been experimentally demonstrated \cite{Kwiat:2001aa, Pan:2001aa, Pan:2003aa,Yamamoto:2003aa, Nature_Photonics11.11, Dong:2008aa, doi:10.1126/science.aan0070, PhysRevLett.126.010503}.



The results in Ref. \cite{PhysRevLett.81.5932, PhysRevA.59.169}, however, pointed out that noisy LOCC may not be incorporated into distilling entanglement: ebits cannot be distilled from weakly entangled states. The currently available quantum technologies contain noise \cite{Preskill2018}, which cannot realize entanglement distillation. Despite the importance of entanglement, no result has yet been known to enable entanglement distillation with noisy LOCC.




In this work, we present a protocol for purifying general noisy measurements and resolve the distillation of entanglement with noisy LOCC. The purification protocol incorporates the advantage distillation \cite{256484, 748999} with $(n-1)$ additional qubits on two parties, respectively. We show that the purification protocol with one additional qubit, i.e., $n=2$, is cost-effective for recovering noiseless distillation. We also show that depolarization noise introduces the worst-case scenario that has the smallest range of distillability. In addition, the protocol can be extended to high dimensions. The protocol is robust against measurements and gate errors.
 

\section{Purifying Noisy Measurement}

We begin by presenting the protocol for purifying noisy measurements. It incorporates a cryptographic key agreement protocol called advantage distillation \cite{ 256484, 748999, PhysRevLett.83.4200, PhysRevLett.91.167901, PhysRevA.75.012334, PhysRevA.72.032301}. Throughout, we consider a measurement in the computational basis $M_i = |i\rangle\langle i|$ for $i=0,1$. A noisy measurement may be described by a positive-operator-valued-measure (POVM),
\bea
\widetilde{M}_i = \mathcal{N}[|i\rangle \langle i |] ~~\mathrm{such~that~~} \sum_i \mathcal{N} [|i\rangle \langle i|] =\I. \label{eq:nmn}
\eea
A map satisfying the above is called unital.

In Ref. \cite{PhysRevLett.81.5932, PhysRevA.59.169}, depolarization noise is considered, $D_p (\cdot) = (1-p) (\cdot) + p \frac{\I}{2}$, 
\bea
\widetilde{M}_i = (1-\frac{p}{2})|i\rangle \langle i| + \frac{p}{2} |i\oplus 1\rangle \langle i\oplus 1|  \label{eq:nm}
\eea
where $\oplus$ is a bitwise addition. We have used $\I=|0\rangle\langle 0| + |1\rangle\langle 1|$. A depolarization noise may be regarded as a mixture of all types of noise. In Ref. \cite{PhysRevA.59.169}, it has been shown that measurements with depolarization noise cannot be used to distill entanglement from weakly entangled states, see also Fig. \ref{fig:graph}.

The purification protocol works as follows. We start with general noise in Eq. (\ref{eq:nmn}) and show that depolarization noise corresponds to the worst-case for entanglement distillation. A high-dimensional extension is shown in Supplemental Material. Let $\rho$ denote a qubit state on which a measurement is performed. A noiseless measurement shows probabilities $\tr[\rho M_i]$, and a noisy one finds $\tr[\rho \widetilde{M}_i]$ for $i=0,1$. We here devise a protocol to obtain error-free statistics $\tr[\rho M_i]$ with noisy measurements and additional qubits as resources. 

\begin{figure}[t]
    \centering
    \includegraphics[width=0.3\textwidth]{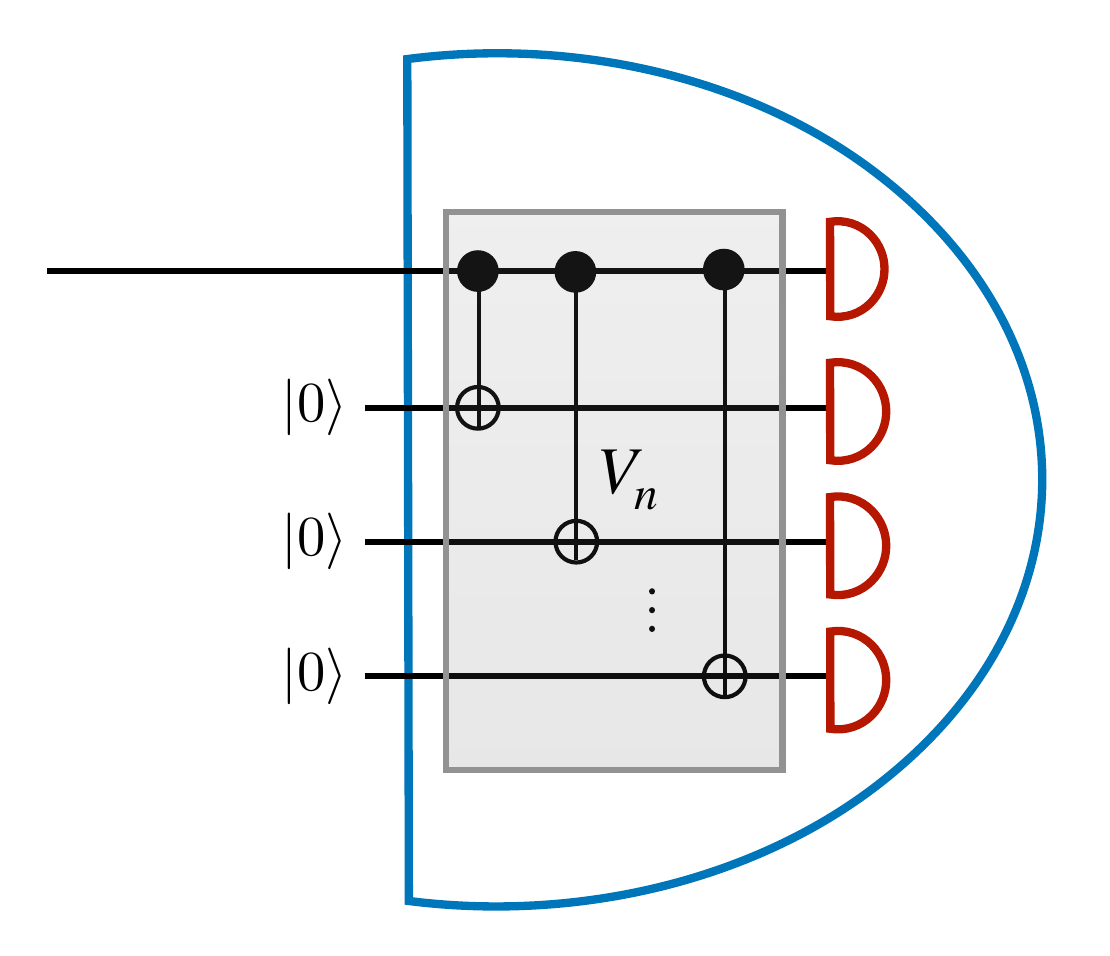}
    \caption{ A measurement is noisy (red), see Eq. (\ref{eq:nm}). An error-suppressing detector (blue) is devised with $n-1$ additional qubits and a collective CNOT $V_n$ in Eq. (\ref{eq:ccnot}).  }
    \label{fig:MEF}
\end{figure}

The protocol requires $(n-1)$ additional qubits and a collective CNOT gate $V_n$ as follows,
\bea
V_n = |0\rangle \langle 0| \otimes \I ^{\otimes n-1} + |1\rangle \langle 1| \otimes X^{\otimes n-1}\label{eq:ccnot}
\eea
with a Pauli $X$ matrix. Then, $(n-1)$ additional qubits are prepared in a fiducial state $|0\rangle^{\otimes n-1}$ and a collective CNOT is applied,
\bea
V_n ~ \rho\otimes |0\rangle \langle 0|^{\otimes n-1} ~V_{n}^{\dagger}. \label{eq:ncnot}
\eea
Measurements on $n$ qubits are noisy. Suppose that $n$-bit outcomes are obtained, $x^n = (x_1, x_2,\cdots, x_n)$ where $x_k \in \{0,1 \}$ for $k=1,\cdots, n$. A POVM element describing the outcomes correspond to $\widetilde{M}_{x^n} = \widetilde{M}_{x_1}\otimes \cdots \otimes \widetilde{M}_{x_n}$. We have the probability of outcome $x^n$,
\bea
p(x^n) = \tr[V_n ~ \rho\otimes |0\rangle \langle 0|^{\otimes n-1} ~V_{n}^{\dagger}   \widetilde{M}_{x^n}]. \label{eq:prob}
\eea
We accept measurement results only when all outcomes are identical, i.e., either $0^n$ or $1^n$; otherwise, discard. This approach has its origin in the advantage distillation \cite{256484,748999,PhysRevLett.83.4200, PhysRevLett.94.020501,PhysRevA.75.012334, PhysRevA.72.032301}. Note also that a similar approach can tolerantly deal with losses in the preparation of cluster states \cite{PhysRevLett.97.120501}. Once outcomes are accepted, a probability of outcome $i \in \{0,1\}$,
\bea
q^{(n)} (i)  = \frac{p(i^n)}{p(0^n ) + p(1^n)} \label{eq:pos}
\eea
which converges to a probability obtained by a noiseless measurement. 

To see that a probability in Eq. (\ref{eq:pos}) is arbitrarily close to a result of a noiseless measurement, let us rewrite Eq. (\ref{eq:prob}) in terms of an effective POVM element for an outcome $x_1\in\{0,1 \}$,
\bea
p(x^n) & = &  \tr [ \rho \widetilde{Q}_{x_1}^{(n)} ],~  \label{eq:prob2} \\
\mathrm{where} & & \widetilde{Q}_{x_1}^{(n)} = \langle 0_2\cdots 0_n | V_{n}^{\dagger} \widetilde{M}_{x^n} V_n | 0_2\cdots 0_n\rangle. \nonumber 
\eea
To see the derivation of Eq. (\ref{eq:prob2}) from Eq. (\ref{eq:prob}), we have used $\tr[AB] = \tr[BA]$ for Hermitian operators $A$ and $B$ so that 
\bea
\tr[V_n ~ \rho\otimes |0\rangle \langle 0|^{\otimes n-1} ~V_{n}^{\dagger}   \widetilde{M}_{x^n}] & = & \tr[~ \rho\otimes |0\rangle \langle 0|^{\otimes n-1} ~V_{n}^{\dagger}   \widetilde{M}_{x^n}V_n ] \nonumber \\
&=& \tr[~ \rho  \langle 0_2\cdots 0_n | V_{n}^{\dagger} \widetilde{M}_{x^n} V_n | 0_2\cdots 0_n \rangle ].  \label{eq:relations} 
\eea
It holds that  $V_n |0_2\cdots 0_n\rangle  =  |0^n\rangle \langle 0| + |1^n\rangle \langle 1|$, from which corresponding POVM elements when the outcomes are identical, $x^n\in\{ 0^n, 1^n\}$, can be found as follows. 

Suppose that as the consequence of unwanted interaction in Eq. (\ref{eq:nmn}), a noisy measurement is given by, 
\bea
   \widetilde{M}_0=
  \left[ {\begin{array}{cc}
   1-s & t \\
   t & s \\
  \end{array} } \right]~~\mathrm{and}
~~
   \widetilde{M}_1=
  \left[ {\begin{array}{cc}
   s & -t \\
   -t & 1-s \\
  \end{array} } \right] \label{eq:nm01}
\eea
where $s\in[0,1/2)$ and $t\in[-\alpha, \alpha]$ with $\alpha= \sqrt{s(1-s)}$ since $\widetilde{M}_i\geq 0$. 
Let us consider cases when $n$ outcomes are identical, i.e., $x_1= \cdots = x_{n} \in \{0,1 \}$. Let $i \in \{0,1 \}$ denote the value accepted in the measurement. We may characterize POVM elements for an outcome $x_1=i \in \{0,1 \}$ in Eq. (\ref{eq:prob}) are given by 
\bea
\widetilde{Q}_{i}^{(n)} & = & r_{0}^{(n)} M_{i} + r_{1}^{(n)} M_{i\oplus 1} +  ((-1)^i t)^n( |0\rangle\langle1|+|1\rangle \langle 0| ) ~~
\nonumber\\
\mathrm{where}&& r_{0}^{(n)} = (1-s )^n~\mathrm{and}~r_{1}^{(n)} = s^n. \label{eq:r}
\eea
The derivations are detailed in Supplemental Material. The probability in Eq. (\ref{eq:pos}) can be rewritten by defining POVM elements $Q_{i}^{(n)}$ on the first qubit when all outcomes are identical:
\bea
q^{(n)} ( i ) &=& \tr[\rho Q_{i}^{(n)} ]~\mathrm{where} ~Q_{i}^{(n)} = \frac{\widetilde{Q}_{i}^{(n)}}{\sum_{i=0,1} p(i^n) } ~ ~~~~\label{eq:puriprob} 
\eea
One can find $n\geq 1$ such that $Q_{i}^{(n)}$ is $\epsilon$-close to a clean one $M_i$;
\bea
\frac{1}{2}\| Q_{i}^{(n)} -  M_i \|_1  \leq \epsilon, \nonumber
\eea
where $\| \cdot \|_1$ denotes a trace norm. i.e., $\| X\|_1 = \tr \sqrt{X^{\dagger}X}$. We stress that for realistic cases, in particular, for depolarizing noise, where we have an error rate less than $10\%$, i.e., $s\leq 0.1$, one additional qubit, i.e., $n=2$, suffices to get $\epsilon = 2\times 10^{-2}$ for which the probability of having identical outcomes is over $0.8$. The analysis is detailed in Supplemental Material.


\section{Distilling Entanglement}

We now consider the distillation of entanglement with noisy local operations. We consider the distillation protocol in Ref. \cite{PhysRevLett.76.722} because of the following reasons. First, it can distill ebits from all two-qubit entangled states. Second, a single CNOT gate by each party suffices. Generalizations of the protocol can be considered, e.g., \cite{CHI2012143, 10.5555/2011763.2011769}, with a greater number of CNOT gates. However, an improvement is not immediate. The generalizations show enhancement in a limited range of an initial fidelity $F$ in Eq. (\ref{eq:fidel}). In addition, a greater number of CNOT gates are experimentally demanding. The protocol in Ref. \cite{PhysRevLett.76.722} is the most robust and experimentally feasible, e.g., \cite{ Pan:2001aa, Pan:2003aa, Nature423, doi:10.1126/science.aan0070}.

Let us summarize the distillation protocol \cite{PhysRevLett.76.722}. It begins with copies of unknown bipartite states $\rho^{\otimes N}$ for a large $N$. Two parties apply state twirling by LOCC, which transforms a shared state to an isotropic state at a single-copy level as follows, 
\bea
\rho \mapsto \rho_F = \sum_{k=1}^m p_k U_k \otimes U_{k}^\ast \rho (U_k \otimes U_{k}^\ast)^{\dagger}\nonumber
\eea
where ${}^\ast$ denotes the complex conjugation with respect to the computational basis.
With unitary transformations that form a quantum $2$-design $\{p_K, U_k \}_{k=1}^m$ \cite{PhysRevA.80.012304, PhysRevA.99.062302}. An isotropic state can be characterized by a single parameter $F$ as follows,
\bea
\rho_F = F | \phi^+\rangle \langle \phi^+| + \frac{1-F}{3} (\I - | \phi^+\rangle \langle \phi^+|),\label{eq:iso}
\eea
where $F = \langle \phi^+ |\rho_F | \phi^+\rangle $ is called a singlet fraction. 

In what follows, the goal is to distill an ebit in Eq. (\ref{eq:ebit}) from copies of isotropic states. In a realistic scenario, one should admit some fraction of errors in achieving an ebit, e.g., one may refer to a scheme for robust self-testing for an ebit, e.g., \cite{PhysRevA.99.052123}. For clarification, let us introduce an $\epsilon$-ball of an ebit
\bea
B_{\epsilon}(|\phi^{+}\rangle) = \{\sigma \in S(\H \otimes \H)~:~ \frac{1}{2}\|\sigma - |\phi^+\rangle \langle \phi^+| \|_1 \leq \epsilon\} \label{eq:ball}
\eea
where $S(\H)$ denotes the set of states on a Hilbert space $\H$. Then, one can distinguish a state $\sigma \in B_{\epsilon} (|\phi^+\rangle)$ and an ideal instance $|\phi^+\rangle$ with a probability at most  $1/2+\epsilon$, which is $\epsilon$ better than the random. From the well-known relation between the state fidelity and the trace distance \cite{761271}, we have for a state $ \sigma \in B_{ \epsilon} (|\phi^+\rangle)$,
\bea
F(\sigma,|\phi^+\rangle) & = & \langle \phi^+| \sigma |\phi^+\rangle \nonumber \\
&=& 1-\frac{1}{2}  \| \sigma - |\phi^+\rangle \langle \phi^+| \|_1 \nonumber
\eea
which is at least $ 1-\epsilon$. Throughout, distilling an ebit signifies a state belonging to $B_{\epsilon}(|\phi^+\rangle)$ admitting a precision up to $\epsilon$ so that we may achieve an ebit with a fidelity higher than $1-\epsilon$.  

\subsection{Noiseless scenario}

\begin{figure}[t]
    \centering
    \includegraphics[width=0.45\textwidth]{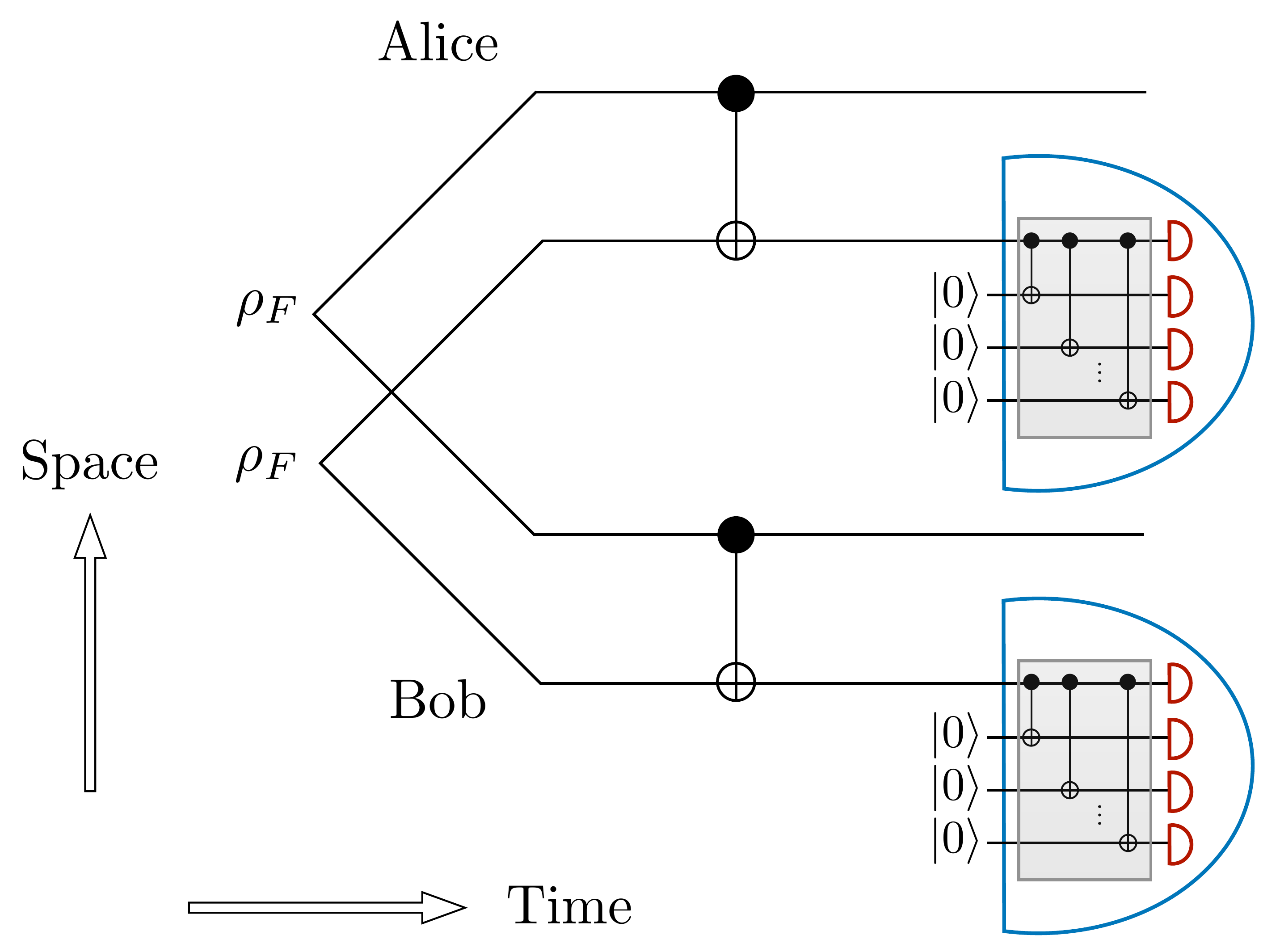}
    \caption{ The protocol for distilling entanglement works with a pair of isotropic states $\rho_{F}^{\otimes 2}$ distributed to Alice and Bob, see Eq. (\ref{eq:iso}). Then, each party applies a CNOT gate and performs a measurement on a qubit in the second register. A resulting state in the first register is accepted only when the measurement outcomes are identical. Then, an ebit can be distilled whenever $F>1/2$. When measurements in the second register are noisy, an ebit can be distilled for a limited range $F>L$ for some lower bound $L>1/2$. Then, two parties apply error-suppressing detectors and recover a lower bound $L$ sufficiently close to $1/2$. For an error rate up to $5\%$, one additional qubit on each party immediately allows entanglement distillation for an initial singlet fraction up to $F>0.505$. }
    \label{fig:mixed state distillation with MEF}
\end{figure}

Let us follow the steps for copies of isotropic states, see also Fig. \ref{fig:mixed state distillation with MEF}. i) Two parties take two copies $\rho_{F}^{A_1B_1} \otimes \rho_{F}^{A_2B_2}$, and apply bilateral CNOT gates, $V_{2}^{A_1A_2}$ and $V_{2}^{B_1B_2}$, see Eq. (\ref{eq:ccnot}). ii) Two parties perform measurements in the second register $A_2B_2$ in the computational basis $\{M_0, M_1 \}$. iii) If outcomes are equal, a state remaining in the first register $A_1B_1$ is accepted. Otherwise, they abort the protocol and start step i) with isotropic states. Once states are accepted, two parties resume the protocol with accepted copies from the state twirling.  

Suppose that measurement outcomes in the second register are identical and the copy in the first register is accepted. The singlet fidelity of the accepted state is given by
\bea
F' =  \frac{F^2 + \left( \frac{1-F}{3} \right)^2}{F^2 + 2 F \left( \frac{1-F}{3} \right) + 5 \left( \frac{1-F}{3} \right)^2}. \label{eq:fidel}
\eea
We have $F'>F$ for $F>1/2$, which is also the condition that a state $\rho_F$ is entangled. Therefore, the protocol above can distill entanglement from all two-qubit entangled states; hence, for two-qubit states, entanglement implies distillability.

\subsection{Noisy scenario}

In Ref. \cite{PhysRevA.59.169}, noisy LOCC are considered in the distillation protocol, particularly, measurements in the second register $A_2B_2$ suffering from a depolarization noise, see Eq. (\ref{eq:nm}). It turns out that the protocol with noisy LOCC cannot distill ebits from all two-qubit entangled states. For instance, for a depolarization noise with $p=0.1$ in Eq. (\ref{eq:nm}), meaning an error rate is $5\%$, entanglement can be distilled only when $F>0.617$ \cite{PhysRevA.59.169}. No entanglement can be distilled if $p>1-1/\sqrt{2}\approx 0.29 $: noise is crucial in the distillation of entanglement.

In what follows, we incorporate the protocol for purifying noisy measurements to the distillation of entanglement, see Fig. \ref{fig:mixed state distillation with MEF}. The goal is primarily to improve a lower bound of an initial fidelity, denoted by $F>L$, such that a resulting fidelity increases $F^{'}>F$. That is, $L=1/2$ for noiseless LOCC and $L>1/2$ for noisy LOCC with a depolarization noise on measurements \cite{PhysRevA.59.169}, and we aim to maintain $L = 1/2$ even if LOCC are noisy. As a byproduct, the improvement also allows us to optimize the resources for entanglement distillation.

For generality, we do not restrict to measurements with a depolarization noise in Eq. (\ref{eq:nm}), considered in Ref. \cite{PhysRevA.59.169}. We begin with measurements containing general noise in Eq. (\ref{eq:nm01}) and seek a lower bound $L^{(n,m)}$ with $n-1$ and $m-1$ target qubits of Alice and Bob, respectively, for purification of noisy measurements.

We now remark that noisy measurements with a depolarization noise in Eq. (\ref{eq:nm}), that is, particular instances with $t=0$ in Eq. (\ref{eq:nm01}), lead to the worst-case scenario. To be precise, let us suppose that an error rate $s$ is given in a noisy measurement in Eq. (\ref{eq:nm01}) and apply them to the distillation of entanglement. A lower bound for the distillability may depend on parameters $s$ and $t$, denoted by $L^{(1,1)} (s,t)$ without the purification protocol yet, turns out to be the highest for $t=0$, which corresponds to a depolarization noise in Eq. (\ref{eq:nm}), see Fig. \ref{fig:graph2}. The proof is detailed in Supplemental Material. Hence, for a given error rate $s$, we have,
\bea
\max_s L^{(1,1)}(s,t) = L^{(1,1)}(s,0),\label{eq:depw} 
\eea
The results in Ref. \cite{PhysRevA.59.169} consider the worst-case scenario.

\begin{figure}[t]
    \centering
    \includegraphics[width=1\textwidth]{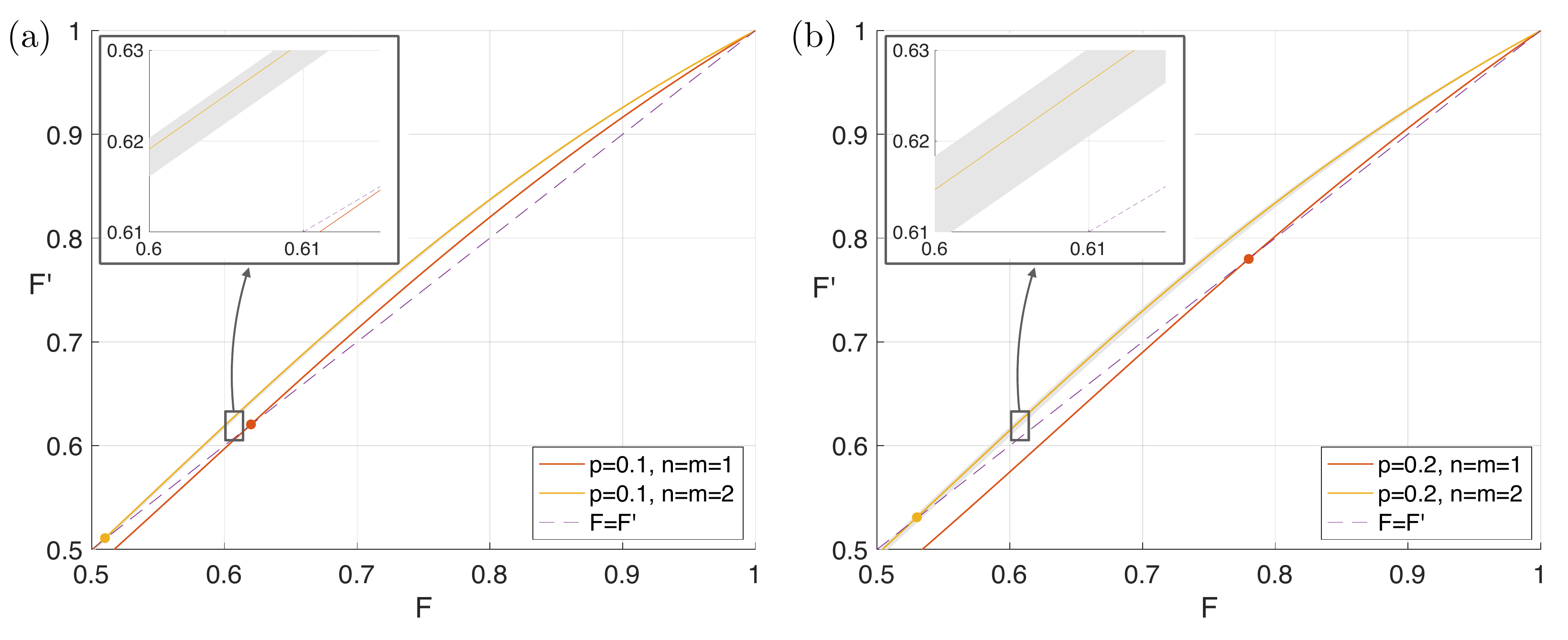}
    \caption{  The robustness of the purification protocol is demonstrated. When a measurement is noiseless, an ebit can be distilled from entangled states having $F>1/2$. (a) For a measurement error $p=0.1$, the condition for distilling entanglement is more stringent $F>0.617$, which is the case $n=m=1$; see the red dot in the plot. With an additional qubit, $n=m=2$, the condition improves up to $F>0.505$; see the yellow dot. Thus, an ebit can be distilled from a larger set of entangled states. (b) For a larger measurement error $p=0.2$, the condition is even worse, $F> 0.779$; see the red dot. With an additional qubit, an ebit can be distilled for $F>0.53$; see the yellow dot. 
    Then, as it is discussed in Eq. (\ref{eq:de}), measurements may have different error rates, for which let $p_i$ denotes an error rate of the $i$-th measurement. The robustness is demonstrated by varying a parameter $\delta_i$ in the range $p_i \in [ p-\delta_i, p+\delta_i  ]$ for $\delta_i \in [0,0.075]$.  With randomly chosen sequences $\delta_i$, the distillation protocol works faithfully to distill ebits; we repeat numerical demonstration and show the range, see the grey area. The robustness also holds for other values of $p$.}
    \label{fig:graph}
\end{figure}

It thus suffices to consider noisy measurements with depolarization noise to show that the purification of noisy measurements improves entanglement distillation.
For a pair of isotropic states obtained after twirling shared states, two parties perform bilateral CNOT gates followed by noisy measurements. Alice and Bob exploit $n-1$ and $m-1$ additional qubits, respectively, for purifying noisy measurements. A resulting singlet fidelity of Alice and Bob relies on measurements on $n$ and $m$ qubits and is computed and expressed as,
\bea
F' = \frac{ F^2 + \left( \frac{1-F}{3} \right)^2 + g^{(n,m)} (F)}{F^2 + 2 F \left( \frac{1-F}{3} \right) + 5 \left( \frac{1-F}{3} \right)^2 +4g^{ ( n,m) } (F)} \label{eq:FF}
\eea
where $g^{(n,m)} (F)$ is defined as follows,
\bea
g^{(n,m)} (F) & = & \left( \frac{ F (1-F) }{3} + \left( \frac{1-F}{3} \right)^2 \right) \frac{ r_{odd}^{ (n,m) } }{ r_{even}^{ (n,m) }}, \label{eq:g1} \\
r_{even}^{(n,m)} & = & (1-\frac{p}{2})^n(1-\frac{p}{2})^m + (\frac{p}{2})^n (\frac{p}{2})^m, ~\mathrm{and}\nonumber \\
r_{odd}^{(n,m)}  & = & (1-\frac{p}{2})^n (\frac{p}{2})^m +  (\frac{p}{2})^n (1-\frac{p}{2})^m. \label{eq:reo}
\eea
For large $n$ and $m$, we have $g^{( n,m)} (F)$ converging to zero and reproduce the fidelity in Eq. (\ref{eq:fidel}) of a noiseless instance. Note that the probability of having identical outcomes in the distillation protocol is given by
\bea
p_{succ}^{(n,m)} (F) &=& r_{even}^{(n,m)}\left(F^2 + 2 F \left( \frac{1-F}{3} \right) + 5 \left( \frac{1-F}{3} \right)^2\right) \nonumber \\
&&  + r_{odd}^{(n,m)} \left(4F\left(\frac{1-F}{3}\right) + 4\left(\frac{1-F}{3}\right)^2\right). \label{eq:succprob}
\eea
In Supplemental Material, we detail in relation to the probability above and the fidelities in Eq. (\ref{eq:FF}), see Table \ref{tab:data}, where one additional qubit is cost-effective.

\begin{figure}[t]
    \centering
    \includegraphics[width=1 \textwidth]{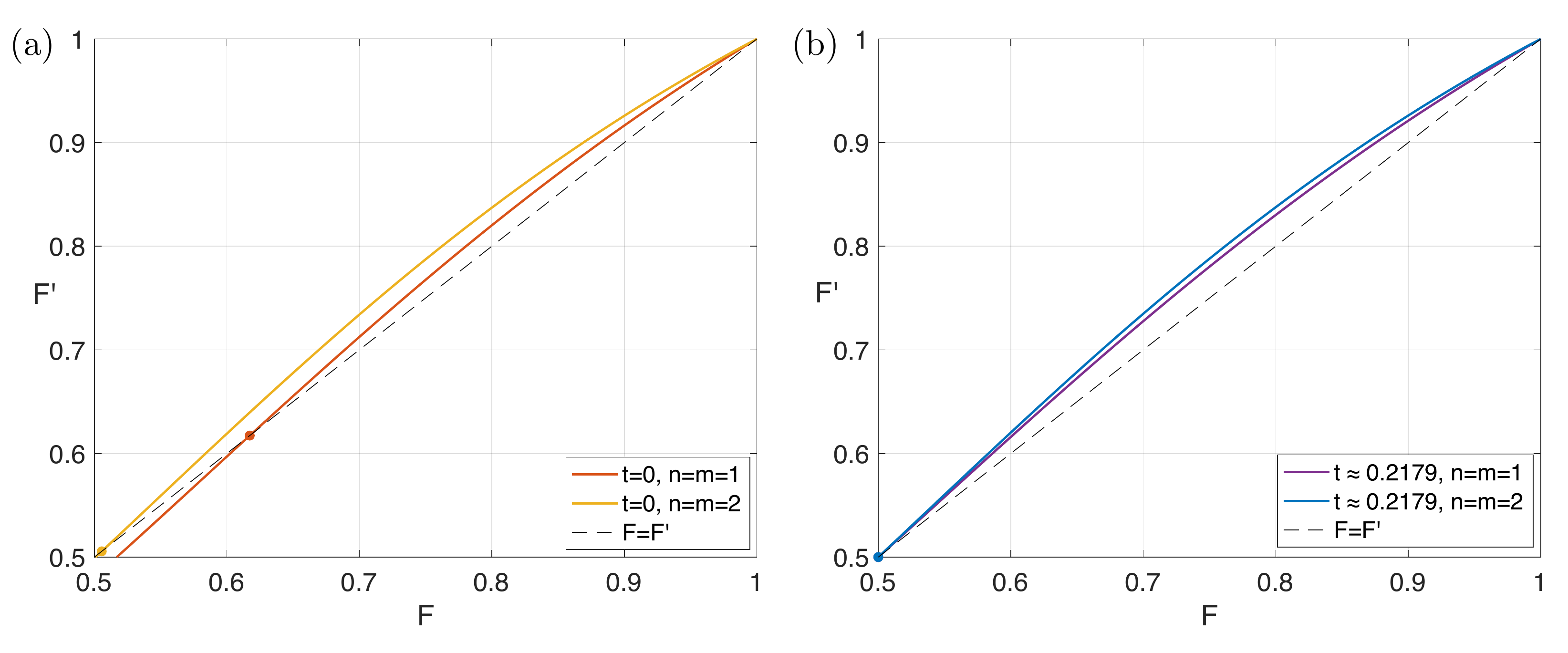}
    \caption{  A measurement error rate, given by $s$, see also Eq. (\ref{eq:nm01}), does not single out a POVM element, which can be characterized by a parameter $t \in [-\alpha,\alpha]$ where $\alpha= \sqrt{s(1-s)}$. For a measurement error $5\%$, i.e., $s=0.05$, we have $\alpha\approx 0.2179$. The worst case that maximizes the lower bound for distillability is when $t=0$, see Appendix C where it is shown that the lower bound decreases as $|t|$ increases. (a) A depolarization noise on measurements, $t=0$, leads to the worst-case scenario in that a lower bound for distillability is the highest. The lower bound for distillability is found $L=0.617$ (red). The purification protocol with a single target qubit ($n=m=2$) improves the lower bound up to $L=0.505$ (yellow). (b) Noisy measurements with $t = \alpha$ in Eq. (\ref{eq:nm01}) show a lower bound for distillability $L=0.5$ (purple). Hence, a noisy measurement can be used to distill entangled states having a fidelity $F>1/2$. In this case, the purification of noisy measurements can improve the rate of distilling entanglement (blue). }
    \label{fig:graph2}
\end{figure}


For the full generality, we suppose that error rates in independent individual measurements are not identical. Let $p_{k}^{A}$ denote a noise fraction of the $k$-th measurement of Alice and $p_{k}^{B}$ for Bob, see Eq. (\ref{eq:nm}). The resulting fidelity is obtained with $g^{(n,m)} (F)$ in Eq. (\ref{eq:g1}) defined by functions $r_{even}^{(n,m)}$ and $r_{odd}^{(n,m)}$,
\bea
r_{even}^{(n,m)} &=& \prod_{i=1}^{n} \left(1-\frac{p^A_i}{2}\right) \prod_{i=1}^{m} \left(1-\frac{p^B_i}{2}\right) + \nonumber \\ &&\prod_{i=1}^{n} \left(\frac{p^A_i}{2}\right) \prod_{i=1}^{m} \left(\frac{p^B_i}{2}\right),~\mathrm{and} \label{eq:de1} 
\eea
\bea
r_{odd}^{(n,m)} &=& \prod_{i=1}^{n} \left(1-\frac{p^A_i}{2}\right) \prod_{i=1}^{m} \left(\frac{p^B_i}{2}\right)  + \nonumber \\ &&\prod_{i=1}^{n} \left(\frac{p^A_i}{2}\right) \prod_{i=1}^{m} \left(1- \frac{p^B_i}{2}\right). \label{eq:de}
\eea
We detail the derivation in Supplemental Material. In Fig. \ref{fig:graph}, we consider an instance with $p_i \in (0.025, 0.175)$ and show that entanglement increases faithfully.

Then, the distillation protocol increases entanglement whenever $F'>F$:
\bea
F'-F &=& - 8\left(F-\frac{1}{4}\right)\left(F- L^{(n,m)} \right)\left(F-1\right) \nonumber \\
\mathrm{where }& & L^{(n,m)} =  \frac{ r_{even}^{ (n,m) }+r_{odd}^{(n,m)}}{2 \left( r_{even}^{ (n,m) }-r_{ odd }^{(n,m)}\right)}.   \label{eq:L}
\eea
We have $F'>F$ for $F\in (L^{(n,m)},1)$ where the lower bound $L^{(n,m)}$ for the distillability depends on the numbers of measurements $n$ and $m$ for purifying measurements and a noise fraction $p$. Note that $L^{(n,m)}$ converges to $1/2$ as $n$ and $m$ tend to be large; hence, the noiseless instance is reproduced.


In a realistic scenario with an error rate $5\%$, we have a cost-effective and resource-efficient strategy $(n,m)=(2,2)$, i.e., single target qubits. A lower bound $L^{(n,m)}$ is computed in Table \ref{tab:LB}.

\begin{table}[t]
\centering
\begin{tabularx}{0.47\textwidth} { 
  | >{\centering\arraybackslash}X 
  | >{\centering\arraybackslash}X 
  | >{\centering\arraybackslash}X 
  | >{\centering\arraybackslash}X 
  | >{\centering\arraybackslash}X 
  | >{\centering\arraybackslash}X 
    | >{\raggedleft\arraybackslash}X |}
 \hline
n & 1 &  2 &  3 &  4  \\
 \hline\hline
$L^{(n,1)}$  & 0.617  &  0.559  & 0.556  & 0.556    \\
\hline
$L^{(n,2)}$ & 0.559  &  0.505  & 0.500  & 0.500   \\
\hline
\end{tabularx}
\caption{The lower bound of fidelity at which entanglement can be distilled. $L^{(n,m)}$ indicates that Alice performs $n$ measurements, while Bob performs $m$ measurements.}
\label{tab:LB}
\end{table}

One can find that the purification of noisy measurements with one additional qubit on each party recovers the noiseless bound $1/2$, see Fig. \ref{fig:graph}.

\subsection{  Robustness} 
 
We detail the analysis when CNOT gates in the purification of noisy measurements are noisy. We consider noise in a collective CNOT gate $V_n$, which may be decomposed as
\bea
V_{n}^{(S_1\cdots S_n)} = \bigotimes_{j=2}^n V_{2}^{(S_1 S_j)}. \label{eq:decom}
\eea
Since a quantum operation is implemented by single-qubit and CNOT operations in practice, we consider a noise fraction $\epsilon$ in a CNOT gate 
\bea
(1- \epsilon) V_2 (~\cdot~) V_{2}^{\dagger} +   \epsilon \frac{\I}{2} \otimes \frac{\I}{2}. \label{eq:noisecnot}
\eea
We here derive a lower bound $L_{\epsilon}^{(n,m)}$ such that $F^{'}>F$ for $F > L_{\epsilon}^{(n,m)}$, where the lower bound relies on an error rate $\epsilon$. In particular, we show that $L_{\epsilon}^{(n,m)}$ reaches an $\epsilon$-close bound to $1/2$. It suffices to consider that $n=m$. A lower bound found is given by,
\bea
L_{\epsilon}^{(n)} & = & \frac{r_{even}^{(n)} + r_{odd}^{(n)}}{2 (r_{even}^{(n)} - r_{odd}^{(n)} ) } = \frac{ (r_{0}^{(n)} + r_{1}^{(n)}   )^2 }{ 2 (r_{0}^{(n)} - r_{1}^{(n)}   )^2 } \nonumber. 
\eea
where $r_{j}^{(n)}$ for $j=0,1$ is a probability obtained by a purified measurement with noise CNOT gates with a noise fraction $\epsilon$. Probabilities $r_{0}^{(n)}$ and $r_{1}^{(n)}$ are obtained by a recurrence relation: 
\bea
r_0^{(n+1)}  & = & (1-\epsilon) r_0^{(n)} (1-\frac{p}{2}) + \frac{\epsilon}{4} (r_0^{(n)} + r_1^{(n)})~~\mathrm{and} ~~ \nonumber \\
r_1^{(n+1)} & = & (1-\epsilon) r_1^{(n)} (\frac{p}{2}) + \frac{\epsilon}{4} (r_0^{(n)}+r_1^{(n)}). \nonumber
\eea
From the relation above, we write by $u$ the limit point, 
\bea
u = \lim_{n\rightarrow \infty} \frac{r_1^{(n )}}{r_0^{(n )}}. 
\eea
It follows that, 
\bea
u= \frac{ (1-\epsilon) u (\frac{p}{2}) + \frac{\epsilon}{4} ( 1+ u ) }{(1-\epsilon)  (1- \frac{p}{2} ) + \frac{\epsilon}{4} ( 1+ u ) },~~\mathrm{from~ which~} ~u = 2(1-p) (1 -\frac{1}{\epsilon}) + \sqrt{5-  4p (2-p) + \frac{4(1-p)^2 }{\epsilon}(\frac{1}{\epsilon} -2) }. \nonumber
\eea
Consequently, when $n$ tends to be large, the lower bound converges $ L_{\epsilon}^{(\infty)}$ as follows,
\bea
 L_{\epsilon}^{(\infty)} = \lim_{n\rightarrow \infty} L_{\epsilon}^{(n)} = \frac{ 2(1-p)^2 (1-2\epsilon) +\epsilon^2 (3 - 2(2-p)p ) + \epsilon \sqrt{ 4(1-p)^2 (1-2\epsilon) +\epsilon^2(5+4 (-2+p)p) } }{4(1-\epsilon)^2 (1-p)^{2} }. \label{eq:conp}
\eea
Note that the lower bound above approaches continuously to $1/2$ as $\epsilon$ converges to $0$.

\begin{figure}[t]
    \centering
\includegraphics[width=0.5\textwidth]{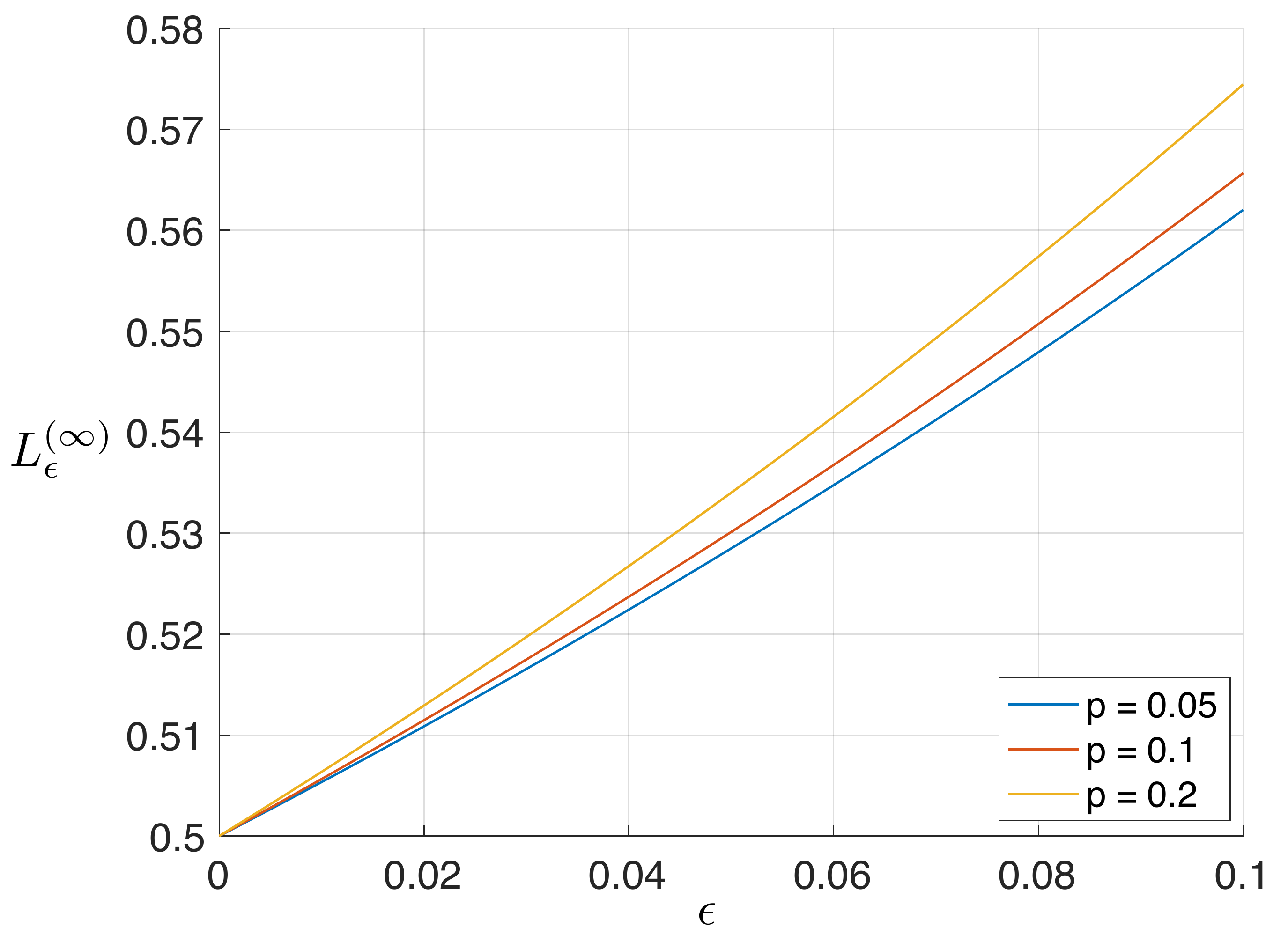}
    \caption{ When a CNOT gate contains an error fraction $\epsilon$, see Eq. (\ref{eq:noisecnot}), the purification of noisy measurements recovers the lower bound for distilling entanglement up to $L_{\epsilon}^{ (\infty )}$ in Eq. (\ref{eq:conp}), which may be strictly larger than $1/2$. The lower bound $L_{\epsilon}^{( \infty)}$ increases as a noise fraction $\epsilon$ increases. }
    \label{fig:image}
\end{figure}

For instance, since errors of a measurement and a CNOT gate are comparable, we consider noise fractions $p=0.1$ in measurements and $\epsilon =0.1$ in the purification protocol and find that one can reach a lower bound $0.566$ with two additional qubits as follows,
\begin{center}
\begin{tabularx}{0.47\textwidth} { 
  | >{\centering\arraybackslash}X 
  | >{\centering\arraybackslash}X 
  | >{\centering\arraybackslash}X 
  | >{\centering\arraybackslash}X 
  | >{\centering\arraybackslash}X 
  | >{\centering\arraybackslash}X 
    | >{\raggedleft\arraybackslash}X |}
 \hline
n & 1 &  2 &  3 &  4  \\
 \hline\hline
$L_{\epsilon}^{(n,n)}$  & 0.617  &  0.570  & 0.566  & 0.566   \\
\hline
\end{tabularx}
\end{center}
where $L_{\epsilon}^{(n,n)}$ denotes the lower bound of an initial fidelity over which entanglement can be distilled. We emphasize that one additional qubit for purification on each party applying entanglement distillation is sufficient for a practical purpose.

\section{ Optimization and Evaluation} 

Having found a cost-effective strategy $(n,m)=(2,2)$ with single additional qubit per party, we investigate the total cost for distilling $1$ ebit. The crucial parameter is the success probability $p_{succ}^{(n,m)} (F)$ in Eq. (\ref{eq:succprob}) for a given fidelity $F$. Note also that a single round of distillation reduces $N$ copies with a fidelity $F$ to $N p_{succ}^{(n,m)}(F)/2$ ones. Let $F_0$ denote an initial fidelity and $F_j$ a fidelity after $j$ rounds. Then, the number of copies $N_c$ to distill $1$ ebit corresponds to
\bea
N_c = \prod_{j=0}^{J-1}   \frac{2}{ p_{succ}^{(n,m)}(F_j) }, \label{eq:ns}
\eea
with $J$ rounds in total, such that $F_J>0.999$, where each round incorporates the purification of noisy measurements with $n-1$ and $m-1$ target qubits.

In Table \ref{tab:data}, the number $N_c$ is computed when noisy measurements have an error rate $5\%$. While a greater number of target qubits can suppress measurement errors, the cost-effective strategy is to have single qubits for purifying noisy measurements $(n,m)=(2,2)$.

\begin{table}[t]
\begin{tabular}{|c|c|c|c||c|}
\hline
$F_0 $ / $ (n,n)$ & (1,1) & (2,2) & (3,3) & Ideal measurement \\
\hline \hline
$  0.6 $ 
& $ L^{(1,1) } >0.6$ & \makecell{ $p_{succ}= 0.498 $ \\$ N_c = 4.6 \times 10^{10}$  } & \makecell{  $ p_{succ} = 0.448$ \\ $ N_c = 1.6 \times 10^{11}$} 
& \makecell{ $p_{succ}= 0.609 $ \\$ N_c = 1.9 \times 10^{8}$  } \\
\hline
$ 0.7 $ 
& \makecell{    $p_{succ} = 0.646$ \\ $N_c = 1.4 \times 10^{12}$} & \makecell{   $p_{succ} = 0.555$ \\ $ N_c = 2.0\times 10^{8}$} & \makecell{  $p_{succ} = 0.500$ \\ $N_c = 5.0 \times 10^{8}$} 
& \makecell{ $p_{succ}= 0.680 $ \\$ N_c = 2.0 \times 10^{6}$  } \\
\hline
$  0.8 $ 
& \makecell{  $p_{succ} = 0.718$ \\ $N_c = 1.9 \times 10^{9}$ } & \makecell{ $p_{succ} = 0.627$ \\ $N_c = 5.2\times 10^{6}$ } & \makecell{ $ p_{succ } = 0.565$ \\ $N_c = 2.6 \times 10^{7}$} 
& \makecell{ $p_{succ}= 0.769 $ \\$ N_c = 1.9 \times 10^{5}$  } \\
\hline
$  0.9$ 
& \makecell{  $ p_{succ} =0.804$ \\ $ N_c = 1.4\times 10^{7}$ } & \makecell{   $  p_{succ} = 0.714$ \\ $N_c = 1.9 \times 10^{5} $ } & \makecell{   $ p_{succ} = 0.644$ \\$ N_c = 7.0\times 10^{5}$} 
& \makecell{ $p_{succ}= 0.876 $ \\$ N_c = 1.3 \times 10^{4}$  } \\
\hline
\end{tabular}
\caption{
The number of copies $N_c$ in Eq. (\ref{eq:ns}) is shown for an initial fidelity $F_0$ when each party exploits $n-1$ target qubits for purifying noisy measurements. We compare scenarios with a 5\% error rate ($p=0.1$) and ideal error-free measurements. The presence of measurement errors not only raises the lower bound of distillable fidelity but also increases the resources needed to distill one ebit. By applying measurement purification, both the lower bound and resource consumption can be improved. Here, $p_{succ}$ denotes the success probability in the first round. Single additional qubit per party $(2,2)$ are cost-effective.}
\label{tab:data}
\end{table}


 \section{Distilling Entanglement from Pure States}

 The purification protocol works when entanglement is distilled from pure states by noisy operations. From the Schmidt decomposition, one can consider a pure state, $\ket{\psi (\theta)}^{(AB)} = \sin \theta \ket{00}^{(AB)} + \cos \theta \ket{11}^{(AB)}$ for $\theta \in (0,\frac{\pi}{4})$. In a noiseless scenario, Bob implements a local operation with Kraus operators, 
\bea
K_0 = \ket{0}\bra{0} + \tan \theta \ket{1}\bra{1}~\mathrm{and}~ K_1 = \sqrt{1 - \tan^2 \theta} \ket{1}\bra{1}. \nonumber
\eea 
which are constructed by measurements on ancillary qubits, $K_{i}^{(B)} =_E\!\!\langle i| U^{(BE)} |0\rangle_E$ for $i=0,1$. Noisy measurements on ancillary qubits cannot distill an ebit. Our protocol for purifying noisy measurements can apply for distilling entanglement. In Supplemental Material, we show the purification in detail. 


\section{Remark}

Two remarks are in order. Firstly, the protocol for purifying noisy measurements can be extended in high dimensions. The generalization is straightforward by exploiting a $d$-dimensional XOR gate \cite{PhysRevLett.82.1056} or a discrete Weyl operator $X_d = \sum_j  | j+1 (\mod d)\rangle \langle j |  $. Secondly, one can also consider an error correction scheme when dealing with measurement outcomes, such as the repetition code. On the one hand, at least two additional qubits are needed for each party to make a majority vote. A larger number of CNOT gates are required, from which a higher chance of having errors is introduced. On the other hand, an error-correction-based scheme is much less efficient. The following is detailed in Supplemental Material. For $p=0.05$, to achieve errors less than $10^{-3}$, it needs four additional qubits on each party, whereas the proposed protocol needs a single additional qubit. Clearly, given additional qubits, our protocol gives a higher fidelity than an error-correction-based one.


\section{Conclusion}

We have shown that a noisy measurement can be purified and applied to the distillation of entanglement. For both cases of mixed and pure states, we have shown entanglement can be distilled by purifying noisy measurements. The purification protocol is robust against measurement and gate errors in implementation. For a realistic scenario where an error rate is about $5\%$, single additional qubits for the purification suffice to recover the noiseless range for distillability: the singlet fraction of $F >1/2$. Our results are a versatile tool to suppress measurement errors for realizing practical entanglement applications with currently available quantum technologies.


\section{Acknowledgements}
The authors thank Markus Grassl and Sibasish Ghosh for helpful discussions and comments. This work is supported by the National Research Foundation of Korea (Grant No. NRF-2021R1A2C2006309, NRF-2022M1A3C2069728), the Institute for
Information \& Communication Technology Promotion (IITP) (RS-2023-00229524), and grant-in-aid of LG Electronics. 

\appendix
\section*{Supplemental Material }
\section{ Purification of noisy POVM elements}

In this section, we show the purification protocol for a general noisy single-qubit measurement. Two POVM elements give a measurement in a noiseless scenario,
\bea
M_0 =
\left[ {\begin{array}{cc}
1 & 0  \\
0 & 0 \\
\end{array} } \right] ~~\mathrm{and}~~
M_1 =
\left[ {\begin{array}{cc}
0 & 0  \\
0 & 1 \\
\end{array} } \right]. 
\eea
A noisy measurement can generally be described by the following POVM elements,
\bea
\widetilde{M}_0=
\left[ {\begin{array}{cc}
1-s & t \\
t & s \\
\end{array} } \right]~~\mathrm{and}
~~
\widetilde{M}_1=
\left[ {\begin{array}{cc}
s & -t \\
-t & 1-s \\
\end{array} } \right] \label{eq:initial}
\eea
where $t\in[-\alpha, \alpha]$ with $\alpha=\sqrt{s(1-s)}$ since a POVM element is non-negative. Thus, the outcome of the measurement above contains noise.

As it is shown in the main text, the purification of noisy measurements aims to have a noiseless outcome in the first arm $x_1$ by exploiting CNOT gates on $n-1$ additional qubits as follows. After CNOT gates for $n-1$ qubits, we repeatedly apply a noisy measurement. A POVM element giving outcomes $x^n = x_1x_2\cdots x_n$ can be described by
\begin{align}
\widetilde{M}_{x^n} = \widetilde{M}_{x_1}\otimes \cdots \otimes \widetilde{M}_{x_n}. 
\end{align}
We have the probability of outcome $x^n$ is found as follows, 
\bea
p(x^n) = \tr[V_n ~ \rho\otimes |0\rangle \langle 0|^{\otimes n-1} ~V_{n}^{\dagger}   \widetilde{M}_{x^n}]. \label{eq:aprob}
\eea
Measurement results are accepted only when all outcomes are identical, i.e., either $0^n$ or $1^n$; otherwise, discard. Once outcomes are accepted, a probability of outcome $i$, for $i=0,1$, is given by,
\bea
q^{(n)} (i)  = \frac{p(i^n)}{ p_{s}^{(n)}}  ~~\mathrm{where}~~p_{s}^{(n)}=p(0^n ) + p(1^n) \label{eq:spro}
\eea
which may be described by a POVM element given $n-1$ outcomes are identical. 
\bea
p(x^n)  = \tr [ \rho \widetilde{Q}_{x_1}^{(n)} ] ~\mathrm{where} ~ \widetilde{Q}_{x_1}^{(n)} = \langle 0_2\cdots 0_n | V_{n}^{\dagger} \widetilde{M}_{x^n} V_n | 0_2\cdots 0_n\rangle \nonumber 
\eea
for $x_1 = 0,1$. For instance, one can compute the relation, $V_n |0_2\cdots 0_n\rangle  =  |0^n\rangle \langle 0| + |1^n\rangle \langle 1|$. Then, an outcome $x_1$ is accepted when it is identical to the other $n-1$ outcomes; hence, one has $x^n\in\{ 0^n, 1^n\}$. 

One can find a recurrence relation for $\widetilde{Q}_{x_1}^{(k)}$ and $\widetilde{Q}_{x_1}^{(k-1)}$ for $k=2,\cdots,n$:
\bea
\widetilde{Q}_{x_1}^{(k)} &= \bracket{0_{k}}{ ~ V_{2}^{\dagger} ~ \widetilde{Q}_{x_1}^{(k-1)} \otimes \widetilde{M}_{x_k} V_2~}{0_{k}} 
\label{QRecursive}
\eea
where $x_1=x_k$. An initial condition is given in Eq. (\ref{eq:initial}): $\widetilde{Q}_{0}^{(1)} = \widetilde{M}_0$ and $\widetilde{Q}_{1}^{(1)} = \widetilde{M}_1$. 
Solving the recurrence relation, we have 
\bea
\widetilde{Q}_{0}^{(n)}=
\left[ {\begin{array}{cc}
(1-s)^n & t^n \\
t^n & s^n \\
\end{array} } \right]~~\mathrm{and}
~~
\widetilde{Q}_{1}^{(n)}=
\left[ {\begin{array}{cc}
s^n & (-t)^n \\
(-t)^n & (1-s)^n \\
\end{array} } \right] \label{eq:ninitial}
\eea
or express them as, 
\bea
\widetilde{Q}_{i}^{(n)} & = &  (1-s)^n M_{i} + s^n M_{i\oplus 1} +  ((-1)^i t)^n( |0\rangle\langle1|+|1\rangle \langle 0| )
\eea
for $i=0,1$. Once outcomes are accepted, corresponding POVM elements for outcomes $i^n$ can be described as
\bea
 {Q}_{i}^{(n)} = 
 \frac{1}{ p(0^n) + p(1^n) }  \widetilde{Q}_{i}^{(n)}. \label{eq:pq}
\eea

The accept probability for a state $\rho$ can be found as 
\begin{align*}
p(0^n) + p(1^n) &= \tr[\rho(\widetilde{Q}_{0}^{(n)}+\widetilde{Q}_{1}^{(n)})] \\
&= \tr[\rho (((1-s)^n+s^n)\I + (t^n + (-t)^n)X] \\
&= (1-s)^n+s^n + (t^n + (-t)^n) \tr[\rho X] \\
\end{align*}
Note that $t^n + (-t)^n$ vanishes for odd $n$. For even $n$, we have $|t^n + (-t)^n|= 2|t^n| \le 2\sqrt{(1-s)^n s^n}$ from the fact that $|t| \le \sqrt{(1-s)s}$. Therefore we have a lower and upper bound of the accept probability as
\begin{align*}
((1-s)^{\frac{n}{2}}-s^{\frac{n}{2}})^2 \le p(0^n) + p(1^n) \le ((1-s)^{\frac{n}{2}}+s^{\frac{n}{2}})^2.
\end{align*}

\begin{table}[h]
\begin{tabular}{|c|c|c||c|c||c|c|}
\hline
error rate($s$)/ purification(n) & $~~D^{(1)}~~$ & $~~p_{s}^{(1)}~~$ & $~~D^{(2)}~~$ & $~~p_{s}^{(2)}~~$ & $~~ D^{(3)}~~$ & $~~p_{s}^{(3)}~~$ \\
\hline \hline
$s=0.25$ & \makecell{$0.25$}& \makecell{$1$} 
& \makecell{$0.1$} &  \makecell{$ 0.62$} 
& \makecell{$0.04$  } & \makecell{$0.43$} \\
\hline
$s=0.2 $ &  \makecell{$0.2$} & \makecell{$1$} 
& \makecell{$ 0.06$ } & \makecell{$ 0.68$} 
& \makecell{$ 0.02$ }  & \makecell{$0.52$} \\
\hline
$s=0.15 $ & \makecell{$ 0.15$} & \makecell{$1$} 
& \makecell{$0.04  $ } & \makecell{$0.74$} 
& \makecell{$0.01$}  & \makecell{$0.61$} \\
\hline
$s=0.1 $ & \makecell{$  0.1$} & \makecell{$1$} 
& \makecell{$0.02 $ } & \makecell{$0.82$} 
& \makecell{$0.01$}  & \makecell{$0.73$} \\
\hline
$s=0.05 $ & \makecell{$ 0.05$ } & \makecell{$1$ } 
& \makecell{ $0.01$ } & \makecell{$0.90$} 
& \makecell{$0.01$}  & \makecell{$0.85$} \\
\hline
\end{tabular}
\caption{The purification of a noisy measurement with depolarizing noise having an error rate $s$ by $n-1$ additional qubits is shown. The probability $p_{s}^{(n)}$ in Eq. (\ref{eq:spro}) denotes the probability of having identical outcomes from measurements on $n$ qubits. The distance in Eq. (\ref{eq:pud}), which is also equal to the error rate of a noisy measurement, quantifies how close a purified measurement is to a noiseless one. For $n=1$, without the purification, no post-selection is performed, $p_{s}^{(1)}=1$. It is shown that for realistic cases where an error rate is less than $10\%$, i.e., $s=0.05$ or $s = 0.1$, one additional qubit suffices to achieve an error rate less than $10^{-2}$. In these cases, the success probability $p_{s}^{(2)}$ is over $80\%$. For a higher error rate $s=0.25$, it is remarkable that an error rate is suppressed from $0.25$ to $0.1$ with $1$ additional qubit, having the success probability $p_{s}^{(2)}=0.62$, and then to an error rate $0.04$ with $2$ additional ones, having $p_{s}^{(2)} =0.43$. For realistic cases, it is observed the purification protocol maintains the success probability over $80\%$. For a higher error rate, the success probability does not significantly drop down either. It is worth mentioning the power of one additional qubit by which the purification protocol achieves an error rate in the percent level $10^{-2}$.}
    \label{tab:1}
\end{table}

To quantify the purification, we use the trace distance, $D(A,B) = \frac{1}{2}\| A-B \|_1$ for Hermitian operators $A$ and $B$ where $\| X \|_1 = \tr\sqrt{X^{\dagger} X}$. For $\epsilon > 0$, one can find $n\geq 1$ such that 
\bea
D^{(n)} (Q_{i}^{(n)}) : = D ( {Q}_{i}^{(n)} , M_i ) \leq  \epsilon~~\mathrm{for}~~i=0,1.  \label{eq:pud}
\eea
Thus, noisy measurements can be purified such that they arbitrarily approximate noiseless measurements. One can also find an $n\geq 1$ for an $\epsilon>0$ such that $D(  {Q}_{0}^{(n)} +  {Q}_{1}^{(n)},\I) \leq\epsilon$.

\section{ Derivation of Eq. (19) in the main manuscript}



In this section, we take different error rates in noisy measurements into account by generalizing Eq.(8) in the main manuscript, where all measurements have an identical error rate, to cases where each measurement has a different error probability. In addition, let us assume that a measurement is under a depolarization noise for now. In the next section, we show that a depolarization noise corresponds to the worst-case scenario in the distillation protocol, in particular, in terms of the lower bound for distillability for noisy measurements.

Let $p_i$ denote a noise fraction of the $i$-th measurement, so that
\begin{align}
\widetilde{M}_{x^n} = \widetilde{M}_{x_1}\otimes \cdots \otimes \widetilde{M}_{x_n},~ \text{ where }  ~\widetilde{M}_{x_i} &= (1-\frac{p_i}{2}) M_{x_i} + (\frac{p_i}{2}) M_{x_i \oplus 1}. \label{Mxi}
\end{align}
Recall that a POVM element is found as, 
\begin{align}
\widetilde{Q}_{x}^{(n)} &= \bracket{0_2\cdots0_n}{V_{n}^{\dagger} \widetilde{M}_{x^n} V_n}{0_2\cdots0_n} = r_0^{(n)} M_x + r_1^{(n)} M_{x\oplus1}. \nonumber
\end{align}
In the above, coefficients $r_{0}^{(n)}$ and $r_{1}^{(n)}$ can be obtained recursively from an initial condition $\widetilde{Q}_{x}^{(1)} = \widetilde{M}_{x_1}$, where $r_{0}^{(1)}=(1-\frac{p_1}{2})$ and $r_{1}^{(1)} =\frac{p_1}{2}$, and updating $n$. That is, we exploit the following relation in the update $k=2,\cdots,n$:
\bea
\widetilde{Q}_{x}^{(k)} &= \bracket{0_{k}}{V_{2}^{\dagger} \widetilde{Q}_{x}^{(k-1)} \otimes \widetilde{M}_{x_k} V_2}{0_{k}}, \label{Q recursive}
\eea
where $V_{2}$ is a CNOT gate {that takes the first qubit as a control and the $k$-th qubit as a target.}
From the equations (\ref{Mxi}, \ref{Q recursive}), we have 
\begin{align}
\widetilde{Q}_{x}^{(k)} &= \prod_{i=1}^{k} \left(1-\frac{p_i}{2}\right) \ketbra 0 + \prod_{i=1}^{k} \left(\frac{p_i}{2}\right) \ketbra 1.\nonumber
\end{align}
For generality, let $p_{i}^{R}$ denote a noise fraction of the $i$-th measurement of a party $R\in \{A, B\}$ so that the POVM element above may be written as, 
\bea
r_0^{A} & = & \prod_{i=1}^{n} \left(1-\frac{p_i^A}{2}\right), ~ r_1^{A} = \prod_{i=1}^{n} \left(\frac{p_i^A}{2}\right)~~ \mathrm{and}~\nonumber \\
 ~r_0^{B} &=& \prod_{i=1}^{m} \left(1-\frac{p_i^B}{2}\right), ~ r_1^{B} = \prod_{i=1}^{m} \left(\frac{p_i^B}{2}\right). \label{r01}
 \eea
Two parties then exploit noisy measurements purified by $(n-1)$ and $(m-1)$ additional qubits, respectively, and execute the distillation protocol. Once the protocol is successful, an unnormalized state is obtained,
\begin{align}
p_{succ} \rho' &= F^2 r_{\text{even}} |\phi^+\rangle\langle \phi^+| + F\left(\frac{1-F}{3}\right) \left((r_{\text{even}}+r_{\text{odd}}) \Pi_0 - r_{\text{even}} |\phi^+\rangle\langle \phi^+| \right) \nonumber \\
&\phantom{=} + \left(\frac{1-F}{3}\right)F \left( r_{\text{even}}(\Pi_0 - |\phi^+\rangle\langle \phi^+|) + r_{\text{odd}} \Pi_1 \right) \nonumber \\
&+ \left(\frac{1-F}{3}\right)^2 \left( r_{\text{odd}} \Pi_0 + (2 r_{\text{even}} + r_{\text{odd}})\Pi_1 + r_{\text{even}}|\phi^+\rangle\langle \phi^+| \right), \label{distillation result}
\end{align}
where we have used $\Pi_0 = |00\rangle\langle 00| + |11\rangle\langle 11|$ and $\Pi_1 =|01\rangle\langle 01| + |10\rangle\langle 10|$, and the parameters are given by
\begin{align}
r_{\text{even}} = r_0^{A} r_0^{B} + r_1^{A} r_1^{B}~~\mathrm{and}~~ r_{\text{odd}} = r_0^{A} r_1^{B} + r_1^{A} r_0^{B},
\end{align}
with
\begin{align}
r_{\text{even}}^{(n,m)} &= \prod_{i=1}^{n} \left(1-\frac{p^A_i}{2}\right) \prod_{i=1}^{m} \left (1-\frac{p^B_i}{2}\right) + \prod_{i=1}^{n} \left(\frac{p^A_i}{2}\right) \prod_{i=1}^{m} \left(\frac{p^B_i}{2}\right), \nonumber\\
~\mathrm{and} && \nonumber \\
r_{\text{odd}}^{(n,m)} &= \prod_{i=1}^{n} \left(1-\frac{p^A_i}{2}\right) \prod_{i=1}^{m} \left(\frac{p^B_i}{2}\right) + \prod_{i=1}^{n} \left(\frac{p^A_i}{2}\right) \prod_{i=1}^{m} \left(1-\frac{p^B_i}{2}\right).
\end{align}

 \section{ The worst-case scenario: Depolarization noise}
 
A measurement that contains general noise can be described by POVM elements in Eq. (\ref{eq:initial}). The purification protocol transforms them to POVM elements in Eq. (\ref{eq:ninitial}) where off-diagonal elements do not vanish. Hence, POVM elements containing a depolarization noise may not be the most general consideration. We here investigate noisy POVM elements with $t>0$ and show that a depolarization noise, in fact, introduces the worst-case scenario in the sense that, under the type of noise, a lower bound for distillability of entanglement is the highest; hence, the largest fraction of weakly entangled states are not distillable.

Let us begin with a POVM purified by $n-1$ additional qubits, denoted as follows,
\bea
M_0^{(n)} = \begin{bmatrix}
r_0 & t \\
t & r_1
\end{bmatrix}, ~~\mathrm{and}~~
M_1^{(n)} = \begin{bmatrix}
r_1 & -t \\
-t & r_0
\end{bmatrix}, \label{eq:pmn}
\eea
where $t\in[-\alpha, \alpha]$ with $\alpha=\sqrt{s(1-s)}$ since a POVM element is non-negative.
For convenience, let us also introduce 
\bea
r_{even} & = & r_0^2 + r_1^2, ~~\nonumber \\
r_{odd} & = & 2 r_0  r_1, ~~ \text{and} ~ \nonumber \\
r_{c} & = & 2t^2. \nonumber
\eea
Note that since $M_{0}^{(n)}, M_{1}^{(n)}\geq 0$ we have $0 \le r_{c} \le r_{odd} \le r_{even} \le 1$.

\begin{figure}[t]
    \centering
    \includegraphics[width=1\textwidth]{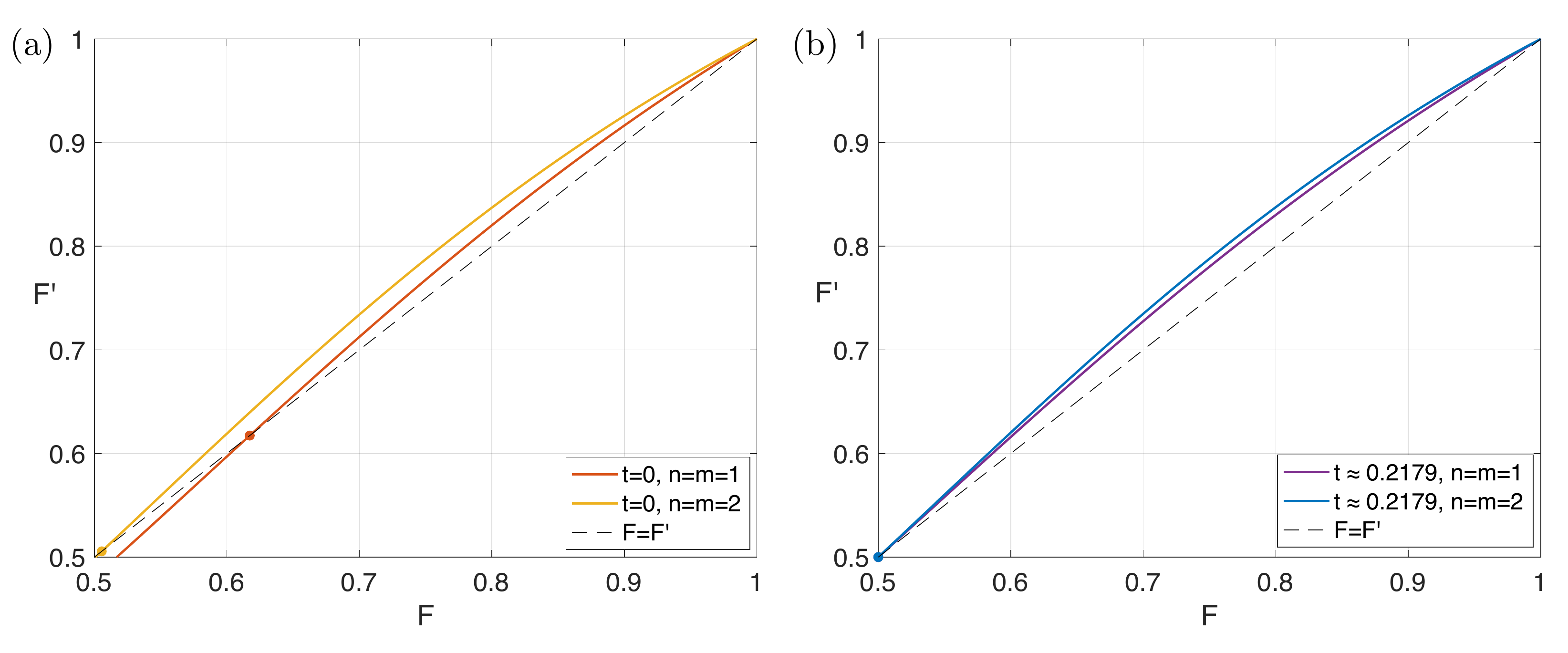}
    \caption{A depolarization noise on measurements leads to the worst-case scenario where a lower bound for distillability is the highest. For an error rate $5\%$, i.e., $s=0.05$ in Eq. (\ref{eq:pmn}), the relation between two fidelities $F^{'}$ and $F$ is shown. (a) Noisy measurements with $t = 0$ in Eq. (\ref{eq:pmn}) provide a lower bound $L=0.617$ for distillation (red). The purification protocol with a single target qubit ($n=m=2$) improves the lower bound up to $L=0.505$ (yellow). (b) Noisy measurements with $t = \alpha\approx 0.2179$ in Eq. (\ref{eq:pmn}) show a lower bound $L=0.5$ (purple). As it is shown in Eqs. (\ref{eq:deri1}) and (\ref{eq:deri2}), the lower bound decreases as $t^2$ increases, meaning that the worst-case is when $t=0$ at which the lower bound is the highest. Note that apart from the lower bound, the purification of noisy measurements can improve the rate of distilling entanglement (blue). }
    \label{fig:graph3}
\end{figure}

Then, the distillation protocol with purified POVM elements in Eq. (\ref{eq:pmn}) applies to copies of shared states having a fidelity $F$. A resulting fidelity is given by 
\begin{equation}
F' = \frac{r_{even} \left( F^2 + \left(\frac{1-F}{3}\right)^2 \right) + r_{odd} \left( F \left(\frac{1-F}{3}\right) + \left(\frac{1-F}{3}\right)^2 \right) + r_{c} \left( F^2 + F \left(\frac{1-F}{3}\right) - 2 \left(\frac{1-F}{3}\right)^2 \right)}{r_{even} \left( F^2 + 2F \left(\frac{1-F}{3}\right) + 5 \left(\frac{1-F}{3}\right)^2 \right) + r_{odd} \left( 4F \left(\frac{1-F}{3}\right) + 4 \left(\frac{1-F}{3}\right)^2 \right) + r_{c} \left( F^2 + 2F \left(\frac{1-F}{3}\right) - 3 \left(\frac{1-F}{3}\right)^2 \right)}. \nonumber
\end{equation}
It turns out that the resulting fidelity relies on not only the diagonal elements $r_{even}$ and $r_{odd}$ but also the off-diagonal elements $r_c$.

One can compute the following:
\bea
\frac{\partial F'}{\partial r_{c}} = \frac{  (1-F) (4F-1) (2F+1) \left(( 7 - 4F)r_{even} + (5 + 4F)r_{odd} \right) }{((5 - 4F + 8F^2)r_{even} + (4 + 4F - 8F^2)r_{odd} + 3(-1 + 4F)r_{c})^2} \label{eq:deri1}
\eea
where it holds that for $F\in [1/2,1]$
\bea
\frac{\partial F'}{\partial r_{c}} \ge 0. \label{eq:deri2}
\eea
Hence, a resulting fidelity $F^{'}$ is monotonically increasing with respect to $r_c$. In other words, a higher $r_c$, i.e., a higher $|t|$, leads to a larger resulting fidelity $F^{'}$. Consequently, a lower bound $L$ is achieved in the distillability condition $F>L$ such that $F^{'}>F$. Thus, a lower bound $L$ has the highest value
\bea 
\max_{r_c} L 
\eea
when $r_c=0$ is the smallest, i.e., $t=0$. This shows that, given an error rate $s$, a depolarization noise in a measurement introduces the worst-case, meaning that a lower bound for distillability is the highest. We also remark that the consideration in Ref. \cite{PhysRevA.59.169} corresponds to the case where, given a measurement error rate, the lower bound for the distillability is the highest.\\

We remark that for all values of $t$ in Eq. (\ref{eq:pmn}) such that $M_{i}^{(n)}\geq 0$ for $i=0,1$, apart from the worst-case scenario in terms of a lower bound for distillability, the purification of noisy measurements also improves the rate of distilling entanglement. For instance, as it is shown in Fig. \ref{fig:graph3}, noisy measurements with $t\neq 0$ have a lower bound closer to $1/2$ than the case with $t=0$, and the purification of measurements then enhances the rate of distilling entanglement, see the comparison of plots in purple and blue in Fig. \ref{fig:graph3}.

\section{  Optimization and Evaluation: Resources for Distilling Entanglement }

In this section, we investigate the success probability of the distillation protocol equipped with the purification of noisy measurement with $(n-1)$ and $(m-1)$ additional qubits on two parties, respectively. As it is shown in the main text, a resulting singlet fidelity of Alice and Bob relies on measurements on $n$ and $m$ qubits. The distillation protocol obtains a resulting fidelity $F'$ from an initial fidelity $F$.

\bea
F' = \frac{ F^2 + \left( \frac{1-F}{3} \right)^2 + g^{(n,m)} (F)}{F^2 + 2 F \left( \frac{1-F}{3} \right) + 5 \left( \frac{1-F}{3} \right)^2 +4g^{ ( n,m) } (F)} \label{eq:FF1}
\eea
where $g^{(n,m)} (F)$ is defined as follows,
\bea
g^{(n,m)} (F) & = & \left( \frac{ F (1-F) }{3} + \left( \frac{1-F}{3} \right)^2 \right) \frac{ r_{odd}^{ (n,m) } }{ r_{even}^{ (n,m) }}, \label{eq:g} \\
r_{even}^{(n,m)} & = & (1-\frac{p}{2})^n(1-\frac{p}{2})^m + (\frac{p}{2})^n (\frac{p}{2})^m, ~\mathrm{and}\nonumber \\
r_{odd}^{(n,m)}  & = & (1-\frac{p}{2})^n (\frac{p}{2})^m +  (\frac{p}{2})^n (1-\frac{p}{2})^m. \label{eq:reo}
\eea
As $n$ and $m$ tend to be large, we have that $g^{( n,m)} (F)$ converges to zero and reproduce the noiseless case. 

The success probability, i.e., the probability that the copy in the first register is accepted in the distillation protocol, is given by
\bea
p_{succ}^{(n,m)}(F) = \left( F^2 + 2F \left(\frac{1-F}{3}\right) + 5\left(\frac{1-F}{3}\right)^2 \right) r_{even}^{(n,m)} 
+ \left(4F \left(\frac{1-F}{3}\right) + 4F \left(\frac{1-F}{3}\right)^2 \right) r_{odd}^{(n,m)}. \label{eq:psucc}
\eea
With the success probability, the resulting fidelity in Eq. (\ref{eq:FF1} ) can be rewritten as follows,
\begin{align*}
F' = \frac{1}{p_{succ}^{(n,m)}(F)} \left[
\left( F^2 + \left(\frac{1-F}{3}\right)^2 \right) r_{even}^{(n,m)}
+ \left( F\left(\frac{1-F}{3}\right) + \left(\frac{1-F}{3}\right)^2 \right) r_{odd}^{(n,m)}
\right]. 
\end{align*} Note that the success probability $p_{succ}^{(n,m)}(F)$ relies on two parameters, the number of target qubits $n-1$ and $m-1$ and an initial fidelity $F$.

We emphasize that the success probability in Eq. (\ref{eq:psucc}) takes into account two steps: one is the probability that the purification protocol with $n-1$ or $m-1$ qubits is successful, and the other is the probability that the first copy is accepted in the distillation protocol. Then, two strategies may be considered.

The first is to focus on the purification of noisy measurements. Once the purification is successful, i.e., with a smaller error rate in measurements, the probability of accepting the first copy in the entanglement distillation protocol gets higher. Then, $n-1$ and $m-1$ qubits are applied with relatively larger $n$ and $m$, and the probability of having identical outcomes may become lower, see Table \ref{tab:1}.

The other is to maintain a larger probability in the purification, i.e., no additional qubit is exploited to purify noisy measurements, and run the distillation protocol with noisy measurements. The probability of accepting the first copy in the distillation protocol becomes lower because of noisy measurements.

We investigate the resources for distilling $1$ ebit with noisy measurements via the purification of noisy measurements with $n-1$ and $m-1$ additional qubits. The constraint is put to achieve a fidelity with an ebit higher than $0.999$. 
Note that each round of the distillation protocol makes a post-selection of $N p_{succ}^{(n,m)}/2$ copies out of $N$ shared copies on average. Let $F_j$ denote a singlet fidelity after $j$ rounds: an initial fidelity is thus denoted by $F_0$ and a final one $F_J >0.999$.

Then, the critical number of copies $N_c$ having an initial fidelity $F_0$ can be computed for distilling $1$ ebit,
\begin{align*}
N_c = \prod_{j=0}^{J-1}   \frac{2}{ p_{succ}^{(n,m)}(F_j) }, 
\end{align*}
with the purification of noisy measurements with $n-1$ and $m-1$ qubits. The results for realistic scenarios where a measurement error is $5\%$ are shown in Table \ref{tab:fidelity}. It is shown that one additional qubit on each party for purifying noisy measurements is cost-effective for distilling entanglement. 

\begin{table}[h]
\begin{tabular}{|c|c|c|c|c|c|}
\hline
$  $ & \makecell{ $N_c$  \\ $ p_{succ}^{(1,1)} (F_0)$ } &  \makecell{ $N_c$  \\ $ p_{succ}^{(2,2)}(F_0)$ } &  \makecell{ $N_c$  \\ $ p_{succ}^{(3,3)} (F_0)$ } &  \makecell{ $N_c$  \\ $ p_{succ}^{(4,4)} (F_0) $ } &   \makecell{ $N_c$  \\ $ p_{succ}^{(5,5)} (F_0)$ } \\
\hline \hline
$F_0 = 0.6, p=0.1$ 
& undistillable & \makecell{$4.65828\times 10^{10}$ \\ $0.498$} & \makecell{$1.62622\times 10^{11}$ \\ $0.448$} &  \makecell{$1.54521\times 10^{12}$ \\ $0.404$} & \makecell{$1.47579\times 10^{13}$ \\ $0.365$} \\
\hline
$F_0 = 0.7, p=0.1$ & \makecell{$1.4316\times 10^{12}$ \\ $0.646$} & \makecell{$2.01242\times 10^{8}$ \\ $0.555$} & \makecell{$5.00584\times 10^{8}$ \\ $0.500$} & \makecell{$3.16687\times 10^{9}$ \\ $0.451$} & \makecell{$2.00696\times 10^{10}$ \\ $0.407$}  \\
\hline
$F_0 = 0.8, p=0.1$ & \makecell{$1.93742\times 10^{9}$ \\ $0.718$} & \makecell{$5.19487\times 10^{6}$ \\ $0.627 $} & \makecell{$2.64262\times 10^{7}$ \\ $0.565 $} & \makecell{$1.36316\times 10^{8}$ \\ $0.510 $} & \makecell{$7.03678\times 10^{8}$ \\ $0.460 $} \\
\hline
$F_0 = 0.9, p=0.1$ & \makecell{$1.42925\times 10^{7}$ \\ $0.804 $} & \makecell{$1.86684\times 10^5$ \\ $0.713 $} & \makecell{$7.04858\times 10^5$ \\ $0.643 $} & \makecell{$2.67407\times 10^{6}$ \\ $0.581$} & \makecell{$1.01473\times 10^{7}$ \\ $0.524 $} \\
\hline
\end{tabular}
\caption{The number of copies of shared states $N_c$ for distilling an ebit is computed when a measurement error is $5\%$. The success probability of the first round starting from $F_0$ with $n-1$ and $m-1$ additional qubits on parties for the purification is shown, $p_{succ}^{(n,n)}$. It is shown that cases with one additional qubit per party, i.e., $(2,2)$, are the most efficient in distilling $1$ ebit. The success probability does not significantly drop down. For the case $(2,2)$, the first success probability with $n$ qubits on each party for purifying noisy measurements: $p_{succ}^{(2,2)} (F_0)$ is comparable to $p_{succ}^{(1,1)} (F_0)$. }
    \label{tab:fidelity}
\end{table}

\section{Advantage Distillation versus Repetition Code}

The proposed protocol for purifying noisy measurements exploits $n-1$ qubits. Noisy measurements are performed on $n$ qubits repeatedly and the outcomes are accepted only when $n$ outcomes are identical, either $0^n$ or $1^n$, which has the origin in the advantage distillation from the cryptographic protocol \cite{256484, 748999}. This has been exploited for linking quantum and classical key distillation protocols \cite{PhysRevLett.83.4200, PhysRevLett.91.167901, PhysRevA.75.012334, PhysRevA.72.032301}. 

Then, it is shown that the purification protocol for noisy measurements with $n-1$ additional qubits introduces a POVM element in post-selection, see also the derivation in Eq. (\ref{eq:pq})
\bea
Q_{i}^{(n)}  = \frac{ (1-\frac{p}{2})^n  }{p(0^n) + p(1^n) } M_i + \frac{ ( \frac{p}{2})^n  }{p(0^n) + p(1^n) } M_{i\oplus 1}, \label{eq:ad}
\eea
where a depolarization noise, introducing the worst-case scenario, is considered, i.e., $t=0$.

One can also consider a strategy of majority vote instead of advantage distillation. For measurements on $n=2m+1$ qubits, let $m_i$ denote the number of $i$ for $i=0,1$, i.e., $m_0 + m_1 = 2m+1$. Then, one may conclude $0$ if $m_0>m_1$. Or, one makes a conclusion $1$ for $m_1>m_0$. This is precisely a decoding process in the repetition code $R_{2m+1}$. One can assert the usefulness that no post-selection is executed. 

In fact, a POVM element can be purified by using the strategy of majority vote and can be obtained as follows,
\bea
R_{i}^{(2m+1)} = \left( \sum_{k=m+1}^{2m+1} \binom{2m+1}{k} (1-\frac{p}{2})^{k} (\frac{p}{2})^{2m+1-k} \right) M_i + \left(\sum_{k=0}^{m} \binom{2m+1}{k} (1-\frac{p}{2})^{k} (\frac{p}{2})^{2m+1-k} \right) M_{i \oplus 1}. \label{eq:rn}
\eea
From the law of large numbers, one can also make an approximate to a noiseless POVM element $M_i$ with $n-1=2m$ additional qubits such that for a given $\epsilon>0$, see also Eq. (\ref{eq:pud})
\bea
D( R_{i}^{(2m+1)}, M_i ) \leq \epsilon. 
\eea
Hence, it is clear that both strategies of dealing with $n$ measurement outcomes, advantage distillation ($AD$) and a repetition code ($R_n$), can purify noisy measurements. It is worth noting that decoding with $R_n$ does not waste any measurement outcomes, while advantage distillation is a post-selection scheme. 

\begin{table}[t]
\centering
\begin{minipage}{.45\textwidth}
\centering
\begin{tabular}{|c|c|c|}
\hline
\multicolumn{3}{|c|}{Repetition code ($R_n$)} \\
\hline
$~n~$ & Fidelity ($1-D$) & Success probability \\
\hline
1 & 0.975 & 1 \\
\hline
3 & 0.99522 & 1 \\
\hline
5 & 0.99921 & 1 \\
\hline
7 & 0.99987 & 1 \\
\hline
\end{tabular}
\subcaption{Repetition code $R_n$ for $p=0.05$}
\end{minipage}%
\hspace{0.5cm} 
\begin{minipage}{.45\textwidth}
\centering
\begin{tabular}{|c|c|c|}
\hline
\multicolumn{3}{|c|}{This work (advantage distillation)} \\
\hline
$~n~$ &  Fidelity ($1-D$) & Success probability  \\
\hline
1 & 0.975 & 1 \\
\hline
2 & 0.99934 & 0.95125 \\
\hline
3 & 0.99998 & 0.92687 \\
\hline
4 & 1 & 0.90369 \\
\hline
\end{tabular}
\subcaption{Advantage distillation for $p=0.05$}
\end{minipage}

\vspace{0.5cm} 

\begin{minipage}{.45\textwidth}
\centering
\begin{tabular}{|c|c|c|}
\hline
\multicolumn{3}{|c|}{Repetition Code ($R_n$)} \\
\hline
$~n~$ &  Fidelity ($1-D$) & Success probability \\
\hline
1 & 0.95 & 1 \\
\hline
3 & 0.98169 & 1 \\
\hline
5 & 0.99424 & 1 \\
\hline
7 & 0.99817 & 1 \\
\hline
\end{tabular}
\subcaption{Repetition code $R_n$ for $p=0.1$}
\end{minipage}%
\hspace{0.5cm} 
\begin{minipage}{.45\textwidth}
\centering
\begin{tabular}{|c|c|c|}
\hline
\multicolumn{3}{|c|}{This work (advantage distillation)} \\
\hline
$~n~$ & Fidelity ($1-D$) & Success probability \\
\hline
1 & 0.95 & 1 \\
\hline
2 & 0.99724 & 0.905 \\
\hline
3 & 0.99985 & 0.8575 \\
\hline
4 & 0.99999 & 0.81451 \\
\hline
\end{tabular}
\subcaption{Advantage distillation for $p=0.1$}
\end{minipage}
\caption{Comparisons between decoding strategies, advantage distillation and majority vote, are shown. For $p=0.05$, in order to achieve an error rate less than $10^{-3}$, the advantage distillation strategy asks a single qubit in addition, that is $n=2$, whereas the majority vote decoding needs $4$ additional qubits, that is, $n=5$. The probability of having two identical outcomes in advantage distillation is over $0.95$. 
For $p=0.1$, in order to achieve an error rate less than $10^{-3}$, the advantage distillation strategy asks two qubits in addition, that is, $n=3$, whereas the majority vote decoding needs more than $6$ additional qubits, that is, $n>7$. The probability of having three identical outcomes in advantage distillation is over $0.85$.}
\label{tab:admv}
\end{table}

In what follows, we investigate comparisons of two strategies. On the one hand, a repetition code is experimentally more demanding. A repetition code asks for a minimal number of qubits, at least $n=3$, for making a majority vote where $2$ additional qubits are necessarily required. However, the proposed purification with advantage distillation is cost-effective for $n=2$, i.e., a single additional qubit. For $n=2$, decoding with a majority vote cannot apply.

On the other hand, the strategy with advantage distillation is more effective in that it achieves a lower error rate than decoding with a majority vote, given the number of additional qubits $n$, it is straightforward to find that
\bea
D(R_{i}^{n},M_i ) > D(Q_{i}^{n},M_i )
\eea
for a given $n$, see also Table \ref{tab:admv} for numerical comparisons $n=2,3, \cdots,7$. For a measurement error rate $5\%$, in a realistic scenario, the success probability in decoding with advantage distillation is over $0.90$ for $n=2$ and $0.85$ for $n=3$. In both cases, we find that the success probability is sufficiently high. In other words, to suppress an error rate up $\epsilon$, decoding with a majority vote needs a strictly larger number of additional qubits than the advantage distillation. Consequently, more CNOT gates are needed. Again, it turns out that decoding with a majority vote is more demanding. After all, we conclude that decoding with a majority vote may be more useful than the advantage distillation when the required number of additional qubits $n$ is sufficiently large, which, however, does not occur in a realistic scenario or even with the currently available quantum technologies.

\begin{figure}[t]
    \centering
    \includegraphics[width=0.45\textwidth]{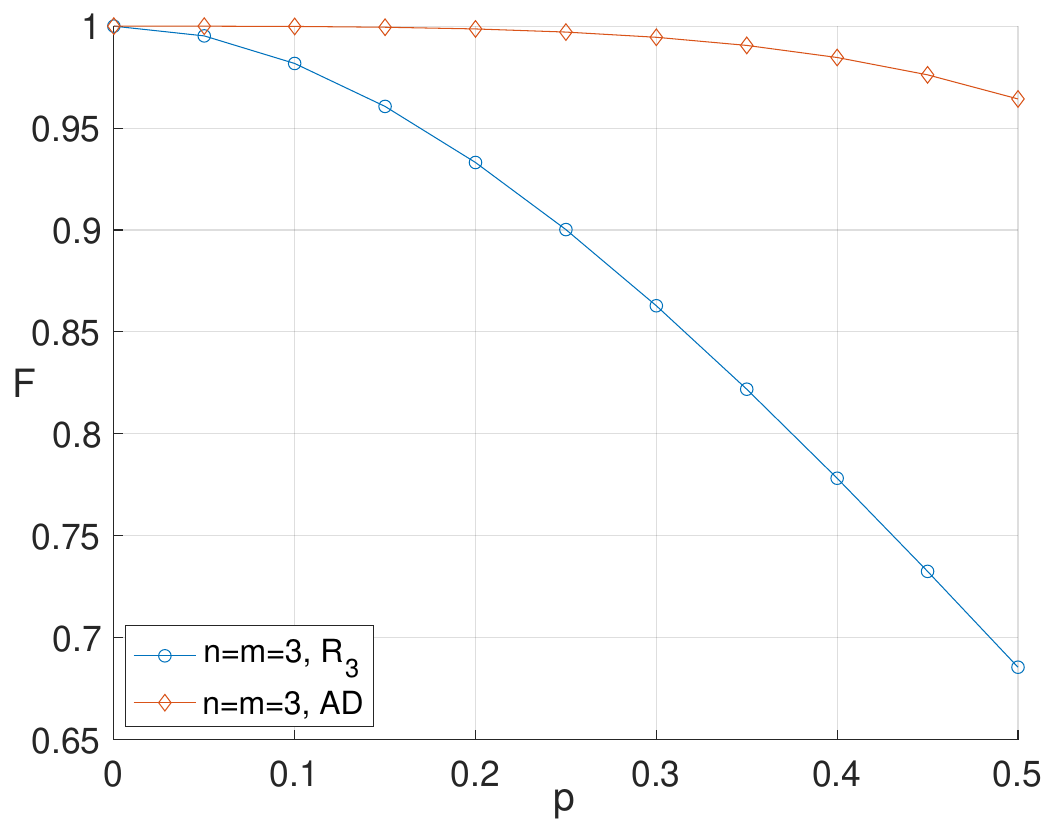}
    \caption{The fidelity $F = 1-D$ where $D$ denotes the trace distance in Eq. (\ref{eq:pq}) is shown for two decoding strategies, advantage distillation (AD) in Eq. (\ref{eq:ad}) and repetition code $R_3$ in Eq. (\ref{eq:rn}) for measurements on $n=3$ qubits. It is shown that an error rate equal to the trace distance $D$ is more effectively suppressed by advantage distillation than the repetition code $R_3$. }
    \label{fig:sp1}
\end{figure}

\section{Distilling entanglement from pure states}

We next consider cases where two parties share pure states which are not maximally entangled. From the Schmidt decomposition, let us suppose that two parties share copies of a state, $\ket{\psi (\theta)}^{(AB)} = \sin \theta \ket{00}^{(AB)} + \cos \theta \ket{11}^{(AB)}$ where $\theta \in (0,\frac{\pi}{4})$. In a noiseless scenario, Bob implements a local operation with Kraus operators, 
\bea
K_0 = \ketbra{0} + \tan \theta \ketbra{1}~\mathrm{and}~ K_1 = \sqrt{1 - \tan^2 \theta} \ketbra{1}. ~~~~~~~ \label{eq:kraus}
\eea
Kraus operators above are realized with the help of an additional system denoted by $E$. Let us introduce a controlled-$W$ gate on systems $BE$
\bea
U^{(BE)} & = &  \ketbra{0}^{(B)} \otimes I^{(E)} + \ketbra{1}^{(B)} \otimes W^{(E)}, ~\mathrm{where} \nonumber \\
W & = &
\begin{pmatrix}
\tan \theta & -\sqrt{1 - \tan^2 \theta} \\
\sqrt{1 - \tan^2 \theta} & \tan \theta
\end{pmatrix}, \label{W}
\eea
such that $W \ket 0 = \tan \theta \ket 0 + \sqrt{1 - \tan^2 \theta} \ket 1$. For a state $\rho^{(BE)}$, a measurement is performed on system $E$ after a controlled-$W$ gate. An outcome $m\in\{0,1 \}$ realizes a Kraus operator $K_m$ in Eq. (\ref{eq:kraus}) on system $B$, i.e., 
\bea
K_{m}^{(B)} =_E\!\!\langle m| U^{(BE)} |0\rangle_E. \label{eq:kraus2}
\eea
When an outcome $0$ occurs and thus $K_0$ is realized, an ebit is distilled. Otherwise, two parties share a product state since the other Kraus operator $K_1$ is rank-one. The probability of having outcome $0$ is given by $2\sin^2\theta$, i.e., we have 
\bea
(\I \otimes K_0) \ket{\psi (\theta)} = \sqrt{2} \sin \theta \ket{\phi^+}. \label{eq:pd} 
\eea
Thus, entanglement can be distilled by local operations with some probability, for which a measurement on a subsystem is essential.

\begin{figure}[t]
    \centering
    \includegraphics[width=0.45\textwidth]{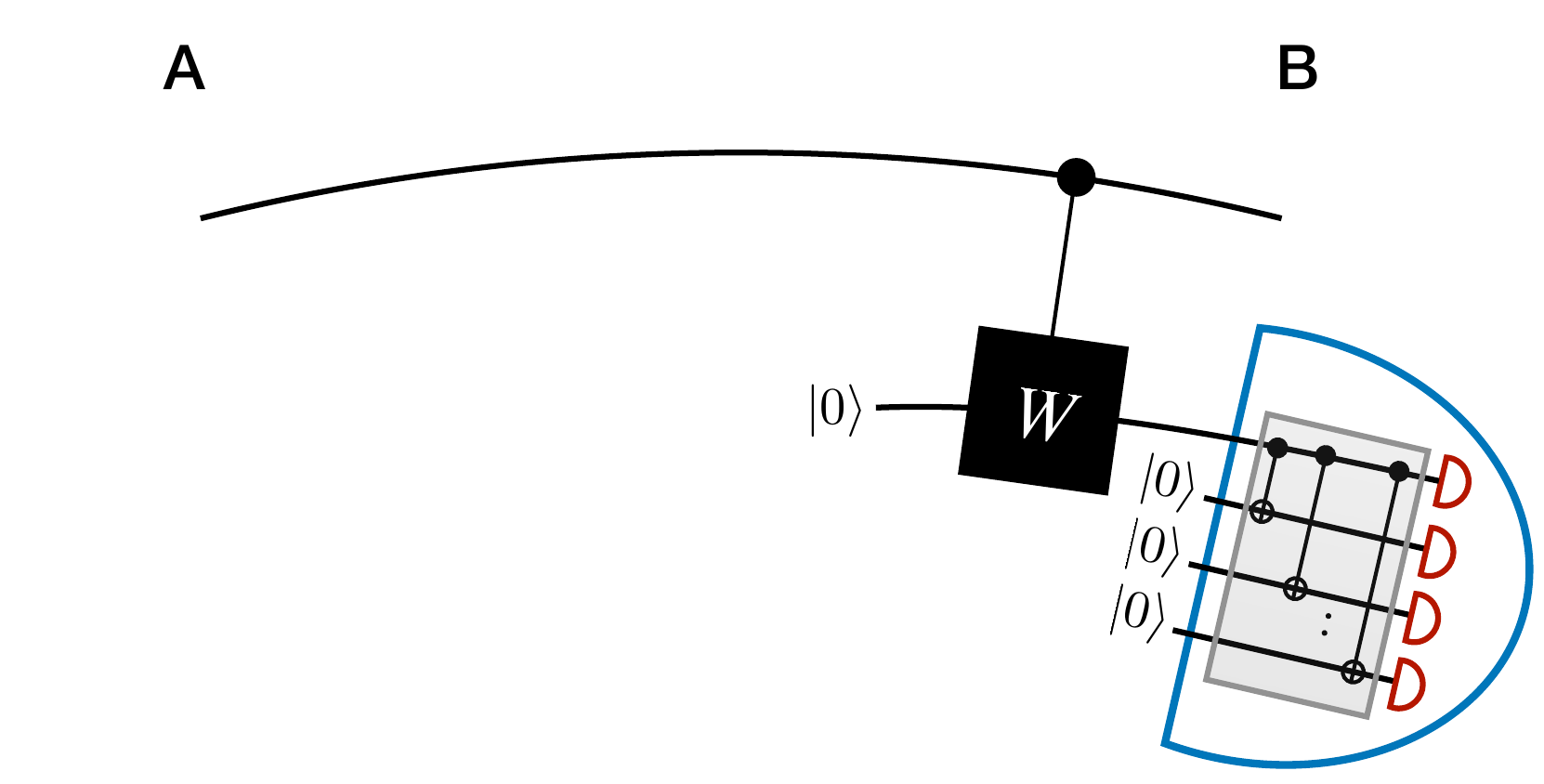}
    \caption{ Two parties sharing a pure state can distill an ebit by applying a local filtering operation on one of the parties, where the local operation is composed of a controlled-$W$ gate in Eq. (\ref{W}) followed by an error-suppressing detector that purifies a noisy measurement.}
    \label{fig:pure state distillation with MEF}
\end{figure}

If a measurement is noisy, Kraus operators in Eq. (\ref{eq:kraus2}) cannot be precisely realized, and consequently, entanglement is not distilled. In particular, two parties may share mixed states due to noise existing in Kraus operators. Similarly to Eq. (\ref{eq:pd}), we have that
\bea
 && \tr_E [U^{(BE)} |\psi\rangle \langle \psi|^{(AB)} \otimes |0\rangle \langle 0|^{(E)}U^{(BE)\dagger}  \widetilde{M}_{0}^{(E)} ] = \label{eq:pbn} \\
 && (1-\frac{p}{2}) K_{0}^{(B)} |\psi \rangle \langle \psi |^{(AB)} K_{0}^{(B) \dagger} +   \frac{p}{2}   K_{1}^{(B)} |\psi\rangle\langle \psi |^{(AB)} K_{1}^{(B) \dagger} \nonumber
\eea
where a measurement $\widetilde{M}_{0}^{(E)}$ is noisy. Thus, a Kraus operator in Eq. (\ref{eq:kraus2}) when a measurement is noiseless no longer applies. Even if a measurement outcome $0$ occurs, the other rank-one Kraus operator $K_1$ appears with a probability $p/2$. 

Purification of a noisy measurement applies to $\widetilde{M}_{0}$ in Eq. (\ref{eq:pbn}) with $n-1$ additional qubits, and let $\rho_{n}^{(AB)}$ denote the resulting state after Bob's operation when $n$ measurement outcomes are identical. Once a resulting state is accepted, the fidelity is given by,
\bea
F_{n} &=& \langle \phi^+| \rho_{n}^{(AB)} | \phi^+\rangle  \nonumber\\
&=& \frac{1}{p_{succ}} \left( {(1-\frac{p}{2})^n 2 \sin^2 \theta  + \frac{1}{2} (\frac{p}{2})^n (1-2 \sin^2 \theta)} \right) \nonumber
\eea
where $p_{succ}$ is the probability of $n$ identical outcomes in the purification protocol, 
\bea
p_{succ} = (1-\frac{p}{2})^n 2 \sin^2 \theta  +  ( \frac{p}{2} )^n (1-2 \sin^2 \theta).  \nonumber
\eea
As $n$ tends to be large, the fidelity converges to $1$, i.e.,
\bea
F_n \rightarrow 1~~\mathrm{as} ~~ n \rightarrow \infty.  \nonumber
\eea
Thus, we have shown distillation of an ebit from a pure state by purifying a noisy measurement. In fact, $F_n$ converges to $1$ as $n$ increases, e.g., $p=0.1$ and $\theta=\pi / 16$: 

\begin{center}
   \begin{tabularx}{0.47\textwidth} { 
  | >{\centering\arraybackslash}X 
  | >{\centering\arraybackslash}X 
  | >{\centering\arraybackslash}X 
  | >{\centering\arraybackslash}X 
  | >{\centering\arraybackslash}X 
   | >{\centering\arraybackslash}X     
   | >{\raggedleft\arraybackslash}X |}
 \hline
n & 1 &  2 &  3 & 4 \\
 \hline\hline
$F_n$ &  0.805 & 0.984   & 0.999 & 1.000  \\
\hline
\end{tabularx}
\end{center}

\section{ Distilling entanglement from pure states with noisy purification protocols}

Suppose that we realize the purification of a measurement with $(n-1)$ additional qubits to distill entanglement from a single-copy pure state, where a POVM element is denoted by $\widetilde{Q}_{0}^{(n)} = r_0^{(n)} M_0 + r_1^{(n)} M_1$. 
Let $p_{succ}$ denote the probability that the purification of a measurement is successful: we have
\bea
p_{succ}~ \rho_n^{(AB)} & = & \tr_E [U^{(BE)} \ketbra{\psi}^{(AB)} \otimes \ketbra{0}^{(E)}U^{(BE)\dagger}  \widetilde{Q}_{0}^{(E)} ] \nonumber \\
&=& r_0^{(n)} 2 \sin^2 \theta \ketbra{\phi^+} + r_1^{(n)} (1-2 \sin^2 \theta) \ketbra{11}. \nonumber
\eea
In this case, we compute the fidelity as follows, 
\bea
F_n = \langle \phi^+| \rho_{n}^{(AB)} |\phi^+\rangle= \frac{1}{p_{succ}} \left( r_0^{(n)} 2 \sin^2 \theta + \frac{1}{2} r_1^{(n)} (1-2 \sin^2 \theta) \right)~~\mathrm{where}~~p_{succ} &= r_0^{(n)} 2 \sin^2 \theta + r_1^{(n)} (1-2 \sin^2 \theta). \nonumber
\eea

For the robustness of the purification protocol, we consider noisy CNOT gates in the purification of noisy measurements. A collective CNOT gate $V_n$, which may be decomposed as,
\bea
V_{n}^{(S_1\cdots S_n)} = \bigotimes_{j=2}^n V_{2}^{(S_1 S_j)} \nonumber 
\eea
is factorized with CNOT gates. We consider a noise fraction $\epsilon$ in a CNOT gate 
\bea
(1- \epsilon) V_2 (~\cdot~) V_{2}^{\dagger} +   \epsilon \frac{\I}{2} \otimes \frac{\I}{2}. \nonumber 
\eea
For an identical noise fraction $p$, suppose that CNOT gates in measurement purification have an error $\epsilon$. We obtain recurrence relations for $r_0^{(n)}$ and $r_1^{(n)}$:
\begin{align*}
r_0^{(n+1)} &= (1-\epsilon) r_0^{(n)} (1-\frac{p}{2}) + \frac{\epsilon}{4} (r_0^{(n)}+r_1^{(n)})~\mathrm{and} \\
r_1^{(n+1)} &= (1-\epsilon) r_1^{(n)} (\frac{p}{2}) + \frac{\epsilon}{4} (r_0^{(n)}+r_1^{(n)}),
\end{align*}
with a starting condition $r_0^{(1)} = (1-p/2)$ and $r_1^{(1)} = p/2$.

For $\epsilon = 0$, we have $r_0^{(n)} = (1-\frac{p}{2})^n$ and $ r_1^{(n)} = (\frac{p}{2})^n$. One can find that $F_n$ converges to $1$ as $n$ tends to be large in this case. For $\epsilon>0$, we have
\bea
u = \lim_{n\rightarrow \infty} \frac{r_1^{(n )}}{r_0^{(n )}} = 2(1-p) (1 -\frac{1}{\epsilon}) + \sqrt{5-  4p (2-p) + \frac{4(1-p)^2 }{\epsilon}(\frac{1}{\epsilon} -2) } \nonumber
\eea
and
\begin{align*}
\lim_{n\rightarrow \infty} F_n &= \frac{{2 \sin^2 \theta  + \frac{1}{2} u (1-2 \sin^2 \theta)}}{2 \sin^2 \theta  + u (1-2 \sin^2 \theta)} \\
&= \frac{2 \epsilon + \left(-2 + 2 p - 2 \epsilon p + \sqrt{\epsilon^2 (5+4(-2+p)p) + 4(1-p)^2(1-2\epsilon)}\right) \cos(2 \theta)}{2\epsilon + 2\left(-2 + \epsilon + 2 p - 2 \epsilon p + \sqrt{\epsilon^2 (5+4(-2+p)p) + 4(1-p)^2(1-2\epsilon)}\right) \cos(2 \theta)}.    
\end{align*}


For instance, for a noise fraction $p=0.1$, $\epsilon=0.05$ and shared state $\theta=\pi / 16$ we have

\begin{center}
   \begin{tabularx}{0.47\textwidth} { 
  | >{\centering\arraybackslash}X 
  | >{\centering\arraybackslash}X 
  | >{\centering\arraybackslash}X 
  | >{\centering\arraybackslash}X 
  | >{\centering\arraybackslash}X 
   | >{\centering\arraybackslash}X     
   | >{\raggedleft\arraybackslash}X |}
 \hline
n & 1 &  2 &  3 & 4 \\
 \hline\hline
$F_n$ &  0.805 & 0.914   & 0.924 & 0.924  \\
\hline
\end{tabularx}
\end{center}

showing that a fidelity converges to $0.924$ with two additional qubits.

\section{Generalization to high-dimensional systems}

The purification protocol can be extended to noisy measurements for high-dimensional quantum systems. Let $\{M_k =|k\rangle \langle k| \}_{k=0}^{d-1}$ denote a measurement in the computational basis in a dimension $d$. Suppose that a depolarization noise to a measurement is introduced, for $k\in \{0,\cdots,d-1 \}$
\bea
\widetilde{M}_k = (1-p) |k\rangle \langle k| + p \frac{\I}{d} \nonumber
\eea
where a depolarization takes all types of noise with equal weights.

The purification protocol can be generalized to $d$-dimensional noisy measurements with a $d$-dimensional XOR gate \cite{PhysRevLett.82.1056}, denoted by $V^{d}$ that works as follows,
\bea
V^{(d)}|a\rangle |b\rangle = |a\rangle |b\oplus a \rangle ~~\mathrm{where}~~b\oplus a = b+a ~(\mathrm{mod}~ d).\nonumber
\eea
One can also write the gate in the following form,
\bea
V^{(d)} = \sum_{j=0}^{d-1}| j\rangle\langle j| \otimes X_{d}^{j} \nonumber
\eea
where $X_d = \sum_{k=0 }^{d-1} |k+1\rangle \langle k |$ is a bit-translation operator in a dimension $d$, referred to as a discrete Weyl operator. Note that $X_{d}^{0}=\I$ and $X_{d}^{j} = \sum_{k} |k+j\rangle \langle k|$. For our purpose, we introduce a collective XOR gate on $n$ systems, 
\bea
V_{n}^{(d)} = \sum_{j=0}^{d-1}| j\rangle\langle j| \otimes ( X_{d}^{j})^{\otimes n-1}. \nonumber
\eea
The protocol for purifying a noisy measurement $\{\widetilde{M}_k \}_{k=0}^{d-1}$ works as follows. For a $d$-dimensional state $\rho$, we prepare $n-1$ $d$-dimensional ancillary systems in a fiducial state $|0\rangle$,
\bea
V_{n}^{(d)} \rho\otimes |0\rangle \langle 0|^{\otimes (n-1)} V_{n}^{(d)\dagger}. \nonumber
\eea
Noisy measurements are performed on $n$ ancillary qubits. Outcomes are accepted only when $n$ results are identical. Once all outcomes are given by $k$, an effective POVM element $Q_{k}^{(n)}$ can be identified by
\bea
 Q_{k}^{(n)} = \frac{  ( 1- (1- \frac{1}{d})p )^n }{ p_{accept} } M_k + \frac{(\frac{p}{d})^n }{p_{accept}}\sum_{l\neq k} M_l
\eea
where 
\bea
p_{accept} = \sum_{k=0}^{d-1} p(k^n) = ( 1- (1- \frac{1}{d})p )^n   + (d-1)(\frac{p}{d})^n. \nonumber
\eea

Therefore, we have 
\bea
 Q_{k}^{(n)} \rightarrow M_k ~~\mathrm{as}~~n\rightarrow \infty. \nonumber
\eea
We have generalized the protocol for purifying a noisy measurement to dimension $d$.


%

\end{document}